\documentclass[12pt,epsf,showkeys]{article}

\usepackage{graphicx,float}
\usepackage{mathrsfs,array,multirow}
\usepackage{amstext}
\usepackage{subfigure}
\usepackage{bm,latexsym,amsmath,amsfonts,amssymb}
\usepackage[usenames,dvipsnames]{color}
\usepackage{color}
\usepackage[colorlinks=true,linkcolor=blue]{hyperref}
\usepackage{soul}
\usepackage{epsfig}

\newcommand{\be}{\begin{equation}}

\newcommand{\ee}{\end{equation}}

\newcommand{\ba}{\begin{array}}

\newcommand{\ea}{\end{array}}

\begin{document}
\begin{titlepage}
\vspace{.5in}
\begin{flushright}
\end{flushright}
\vspace{0.5cm}

\begin{center}
{\Large\bf Homoclinic orbit and the violation of the chaos bound around a black hole with anisotropic matter fields
 }\\

\vspace{.4in}

  {$\mbox{Soyeon \,\, Jeong}^{\dag}$}\footnote{\it email: tiapflzk@naver.com},\, \,
  {$\mbox{Bum-Hoon \,\, Lee}^{\S\dag}$}\footnote{\it email: bhl@sogang.ac.kr},\,\,
  {$\mbox{Hocheol \,\, Lee}^{\dag}$}\footnote{\it email: insaying@sogang.ac.kr},\,\,
  {$\mbox{Wonwoo \,\, Lee}^{\S}$}\footnote{\it email: warrior@sogang.ac.kr}\\

\vspace{.3in}

{\small \S \it Center for Quantum Spacetime, Sogang University, Seoul 04107, Korea}\\
{\small \dag \it Department of Physics, Sogang University, Seoul 04107, Korea}\\

\vspace{.5in}
\end{center}
\begin{center}
{\large\bf Abstract}
\end{center}
\begin{center}
\begin{minipage}{4.75in}

{\small \,\,\,\,
We study the homoclinic orbit and the violation of chaos bound,
which are obtained by particle motions around a black hole that coexist with anisotropic matter fields.
The homoclinic one is associated with an unstable local maximum of the effective potential.
By perturbing a particle located slightly away from the homoclinic one,
we numerically compute Lyapunov exponents indicating the sensitivity of the initial value.
Our results demonstrate that the violation of the chaos bound increases with higher angular momentum,
and the anisotropic matter gives rise to violating the chaos bound further, even in the case of the nonextremal black hole.
We utilize the Hamiltonian-Jacobi formalism to explicitly illustrate how the geodesic motion of a particle
can be integrable in the procedure of obtaining our findings.}
\end{minipage}
\end{center}
\end{titlepage}

\newpage

\section{Introduction \label{sec1}}

\quad

Black holes are not only theoretically predicted in the theory of gravitation, but also become
real celestial objects to exist in the Universe by indirect observations. In the observational aspect, it was
reported that gravitational waves generated by the collision of two black holes were detected~\cite{Abbott:2016nmj,Abbott:2017vtc},
and the existence of black holes was indirectly observed by the detection of light (so-called black hole shadow)
passing through around a supermassive black hole located in the center of a galaxy~\cite{Akiyama:2019cqa, Akiyama:2019bqs, Akiyama:2019fyp, EventHorizonTelescope:2022xnr, EventHorizonTelescope:2022xqj}. Thanks to these observations, a black hole has unveiled its reality.

Studying the geodesic motions and orbits of particles around a black hole is a great way to understand
the geometry around a black hole and the phenomena occurring. The study of these geodesic motions and orbits
by light and massive particles shows that the geometry around a black hole can be
very different from the geometry around a massive object in Newtonian mechanics~\cite{Einstein:1956zz, Bardeen:1973tla, Luminet:1979nyg, Falcke:1999pj, Virbhadra:2002ju, Perlick:2004tq, Virbhadra:2007kw}.

Most motions of objects that occur in nature are not only nonlinear but also could be chaotic~\cite{Barrow:1981sx, Bombelli:1991eg, Dettmann:1994dj}. Actual particle motions described by the theory of gravitation in curved spacetime are nonlinear, and we want to know how chaotic behavior occurs in those motions. In the particle motion around the black hole, the nonchaotic motion of particles is more regular, periodic, and less complicated.
Thus, most of the cases we study are integrable, geodesic, and not chaotic~\cite{Levin:2008yp, Pugliese:2010ps, Pugliese:2011py, Gwak:2008sg, Gwak:2011qs, Uniyal:2014paa, Nandan:2016ksb, Ghaderi:2017wvl, Liu:2017fjx, Al-Badawi:2020htj, Wang:2022ouq}.
In this situation, it is natural to ask how can we know that the particle motions are chaotic around the black hole with a strong gravity effect.
To answer this question, one could explore two kinds of characteristics showing chaos phenomena.
One corresponds to the Poincare section in the phase space, while the other to the Lyapunov exponent~\cite{Cardoso:2008bp, Semerak:2010lzj, Semerak:2012dx, Dalui:2018qqv}.  In this work, we will focus on the Lyapunov exponent and the violation
of chaos bound by the particle motion around the black hole with anisotropic matter fields.

To arrive at this stage, as a nonchaotic motion, we analyze the case corresponding to the homoclinic
orbit~\cite{Levin:2008yp, Perez-Giz:2008ajn} of the particle.
The homoclinic orbit refers to one that approaches two different ones in the infinite future and the infinite past.
This one corresponds to the separatrix between the bound geodesic and the falling geodesic.
The existence of an unstable local maximum of the potential plays the role of giving a separatrix of the particle motion.

We perturb the particle located slightly away from the position of the homoclinic orbit.
One of the consequences of the perturbation in the background of the black hole is that
the geodesic equation loses its complete integrability.
If this integrability is lost, the dynamical system exhibits chaotic
behavior~\cite{Bombelli:1991eg, Vieira:1996zf, Letelier:1997uv, Chen:2003jv, Hashimoto:2016dfz, Liu:2020vsy}.
On the other hand, if the motion of the perturbed particle is integrable again, it may not be the chaotic one.
With the assumption of the particle motion being still geodesic,
we analyze the numerically calculated Lyapunov exponents that indicate
how sensitive this motion of the perturbed particle is to its initial position.

Since the universal upper bound of the Lyapunov exponent defined in thermal quantum systems
was conjectured~\cite{Maldacena:2015waa}, there have been many related studies~\cite{Hosur:2015ylk, Stanford:2015owe,
Rozenbaum:2016mmv, Hashimoto:2020xfr, Yoon:2019cql, Giataganas:2021ghs}.
Here the upper bound is determined by the temperature of the system,
and the study was soon extended to the black hole system.
It has been also studied for cases showing the chaotic behavior of
the spinning particle around the black hole~\cite{Suzuki:1996gm, Kao:2004qs}.
In the black hole system, the temperature can be obtained by the existence of the event horizon,
and it can be anticipated that the upper bound may be violated in an extremal black hole system
where the temperature of the black hole vanishes~\cite{Zhao:2018wkl}.
After that, there have been many studies related to this case in various black hole
geometries~\cite{Lei:2020clg, Mondal:2021exj, Addazi:2021pty, Kan:2021blg, Gwak:2022xje, Gao:2022ybw}.

We study the motion of charged particles around a static black hole that coexists
with matter that has an electric charge and an additional anisotropic matter field~\cite{Kiselev:2002dx, Cho:2017nhx, Visser:2019brz}.
Recently, we noticed a reference one~\cite{Gao:2022ybw},
in which the authors studied chaos bound and its violation in the charged black hole~\cite{Kiselev:2002dx}
with different metric functions compared with our present cases.
In Ref.~\cite{Gao:2022ybw}, the authors focused on the effect of angular momentum,
while we wish to investigate the effect of both anisotropic matter fields and angular momentum in this paper.
When a particle has a charge, there exists an interaction between a particle's one and a black hole's one.
For that reason, the shape of the effective potential that describes the particle's motion could
be modified and then more complicated according to the amount of charge compared with the neutral particle motion.
We investigate how the violation of chaos bound occurs through the values of Lyapunov exponents
when approaching an extremal black hole by adjusting the parameter values in a nonextremal one.
Increasing angular momentum is also known to give rise to the violation of the chaos bound, and we also analyze this one.

This paper is organized as follows: In Sec.\ 2, we show the black hole solutions with anisotropic matter fields.
We analyze the geodesic motion of a particle around a black hole as a homoclinic orbit.
In Sec.\ 3, we perform numerical calculations and analyze the calculated Lyapunov exponents and the chaos bound.
In Sec.\ 4, we summarize and discuss our results.

\section{Homoclinic orbits \label{sec2}}

\quad
In this section, we introduce a well-known black hole solution coexisting with anisotropic matters
and analyze particle motions as the homoclinic orbit around a black hole.

\subsection{Black hole solution}

\quad

We consider the action
\begin{equation}
I=\int d^4x \sqrt{-g}  \Big[\frac{1}{16\pi} (R-F_{\mu\nu}F^{\mu\nu})  +{\cal L}_{\rm am}\Big]   + I_{b}   \,,
\label{action}
\end{equation}
where $R$ is the Ricci scalar   of the spacetime,   $F_{\mu\nu}$ is the electromagnetic field tensor,
we take $G = 1$ for simplicity, ${\cal L}_{\rm am}$ denotes the effective anisotropic matter fields,
and $I_{b}$ corresponds  to the boundary term~\cite{Gibbons:1976ue, Hawking:1995ap}.
We obtain the Einstein equations
\begin{equation}
R_{\mu\nu}-\frac{1}{2}R g_{\mu\nu}=8\pi T_{\mu\nu} \,,
\label{einsteineq}
\end{equation}
where $T_{\mu\nu}=\frac{1}{4\pi}(F_{\mu\alpha}  F_{\nu}^{\alpha}-
\frac{1}{4}g_{\mu\nu}F_{\alpha\beta}F^{\alpha\beta})-
2\frac{{\delta \cal L}_{\rm am} }{\delta g^{\mu\nu}}+ {\cal L}_{\rm am} g_{\mu\nu}$,
and the source-free Maxwell equations
\begin{equation}
\nabla_{\nu}F^{\mu\nu} = \frac{1}{\sqrt{-g}}[\partial_{\nu}(\sqrt{-g}F^{\mu\nu})] =0 \,. \label{maxwelleq}
\end{equation}
We consider a static spherically symmetric black hole solution as the form
\begin{equation}
ds^2 = -f(r)dt^2 + \frac{1}{f(r)}dr^2 + r^2 d\Omega^2_2 \,, \label{metric}
\end{equation}
where we chose the  equation of state is
$p_r=-\rho = -(\rho_e + \rho_{\rm am})$ and
$p_{\theta}=p_{\phi}=\rho_e+ w\rho_{\rm am}$.
The solution is given by~\cite{Kiselev:2002dx, Cho:2017nhx, Visser:2019brz, Kim:2019hfp, Kim:2021vlk}
\begin{equation}
f(r)=1-\frac{2M}{r} +\frac{Q^2}{r^2} - \frac{K}{r^{2w}} \,, \label{metricfunction}
\end{equation}
where   $M$ and $Q$ represent the Arnowitt-Deser-Misner mass and the total charge of the black hole, respectively,
and $K$ is a constant.
The energy density is given by
\begin{equation}
\rho(r) = \rho_e + \rho_{\rm am} =\frac{Q^2}{8\pi r^{4}}+ \frac{r^{2w}_o}{8\pi r^{2w+2}} \,, \label{energydensity}
\end{equation}
where $r_o$ is a chargelike quantity of dimension of length and defined by $r^{2w}_o=(1-2w)K$.
One can see Ref.\ \cite{Cho:2017nhx} for an analysis with $w=1/2$ case.

We now check the classical energy conditions \cite{Cho:2017nhx}.
The energy density is always non-negative when $Q^2 + (1-2w)K r^{2(1-w)}\geq 0$ \cite{Kim:2019hfp}.
And $p_r+\rho=0$, which implies that
the radial null energy condition is satisfied. For the tangential pressures,
the null energy condition is satisfied when $w \geq -1$.
The anisotropic matter fields are those in which the radial and tangential pressures are not equal.
The radial pressure could be negative, and this property allows matter fields or fluids to coexist
with the black hole outside the black hole horizon.
There have been various studies on stars and astrophysical objects
composed of these anisotropic matter fields or fluids~\cite{Bowers:1974tgi, Herrera:1997plx, Mak:2001eb,
Giataganas:2012zy, Kim:2019ojs, Roupas:2020mvs}.

\subsection{Homoclinic orbits}

\quad

To obtain the geodesic equations, we first examine the symmetry and separability structure of the spacetime~\cite{Benenti1979, Demianski1980}.
The relevant quantities correspond to Killing vectors and Killing tensors~\cite{Frolov:2017kze}, and which are known as
$\xi^{\mu}_{(t)}$, $\xi^{\mu}_{(z)}$, $g^{\mu\nu}$ and ${\cal K}^{\mu\nu}$.
They mutually commute under the Schouten-Nijenhuis bracket.
As a result, the geometry admits the separability structure.
Thanks to the spherical symmetry of the geometry, there exist four Killing vectors and two Killing tensors.
Among them, ${\cal K}\equiv {\cal K}^{\mu\nu}p_{\mu}p_{\nu}=p^2_{\theta}+L^2_z/\sin^2\theta$ is equal to $L^2$,
in which $\xi^{\mu}_{(x)}=(0,0, -\cos\phi, \cot\theta\sin\phi)$, $\xi^{\mu}_{(y)}=(0,0, \sin\phi, \cot\theta\cos\phi)$, and
$\xi^{\mu}_{(z)}=(0,0, 0, 1)$~\cite{Walker:1970un}.
Without loss of generality, we restrict the analysis to equatorial orbits $\theta=\pi/2$ thanks to the spherical symmetry of the geometry.
Then ${\cal K}$ becomes $L^2_z$, thus the three quantities $\xi^{\mu}_{(t)}$, $\xi^{\mu}_{(z)}$ and $g^{\mu\nu}$
are sufficient to show the separability structure and obtain the geodesic equations.

We now drive the expressions for a homoclinic orbit as the separatrix.
We consider a test particle with charge $e$ and mass $m$ moving in the black hole geometry (\ref{metricfunction}).

There exit four conserved quantities. Two are related to Killing vectors
\begin{eqnarray}
\xi^{\mu}_{(t)}\pi_{\mu} = -E= -f(r) \frac{dt}{d\lambda} - \frac{eQ}{r} \,, ~~
\xi^{\mu}_{(z)}\pi_{\mu} =L_z= r^2\sin^2\theta \frac{d\phi}{d\lambda}  \,,
\label{killangm}
\end{eqnarray}
where $E$ and $L_z$ correspond to the energy and the angular momentum of the test particle at infinity,
while the other two are related to Killing tensors
\begin{eqnarray}
g^{\mu\nu} p_{\mu}p_{\nu} = -m^2 \,,~~
{\cal K}^{\mu\nu} p_{\mu}p_{\nu} = {\cal K}  = p^2_{\theta} + \frac{L^2_z}{\sin^2\theta } =L^2 \,, \label{constantK}
\end{eqnarray}
where $L$ corresponds to the total angular momentum of the test particle.

We investigate the geodesic motion around the black hole by adopting the Hamilton-Jacobi formalism~\cite{Misner:1974qy}.
The geodesic equations as four first-order differential equations are given by
\begin{eqnarray}
&& r^2 p^t \equiv  r^2 \frac{dt}{d\lambda}  = \frac{P(r)}{f(r)}    \,, \label{HJeq-t}\\
&&r^2 p^{\phi} \equiv  r^2 \frac{d\phi}{d\lambda} =  \frac{L_z}{\sin^2\theta}   \,, \label{HJeq-phi} \\
&& r^2 p^r \equiv  r^2 \frac{dr}{d\lambda} = \pm \sqrt{R(r)}  \,,\label{HJeq-the} \\
&&r^2 p^{\theta} \equiv r^2 \frac{d\theta}{d\lambda} = \pm \sqrt{\Theta(\theta)} \,,  \label{HJeq-theta}
\end{eqnarray}
where signs $+1(-1)$ in Eqs.\ (\ref{HJeq-the}) and (\ref{HJeq-theta}) correspond to the outgoing
(ingoing) geodesics and
\begin{eqnarray}
\Theta(\theta) &=& {\cal Q} +L^2_z  - \frac{L^2_z}{\sin^2\theta} \,, ~~ P(r) = r^2 E -e Q r \,, \nonumber \\
R(r) &=& P^2 - r^2 f(r) [m^2 r^2 + L^2_z + {\cal Q} ] \,, ~~
{\cal Q} = p^2_{\theta} + \frac{\cos^2\theta L^2_z}{\sin^2\theta} \,.  \label{extformr}
\end{eqnarray}
We can take $m \neq 0$ for timelike geodesics and $m = 0$ for null geodesics.

From Eq.\ (\ref{HJeq-the}), we obtain the radial equation for the geodesic motion
\begin{equation}
\frac{1}{2} m \left(\frac{dr}{d \tau} \right)^2  + {\cal V}_{\rm effp} (r)= 0 \,, \label{eqgeo}
\end{equation}
where the effective potential in the equatorial plane is given by
\begin{equation}
{\cal V}_{\rm effp} (r)= \frac{f(r)[m^2 + \frac{L^2_z}{r^2}]-(E-\frac{eQ}{r})^2}{2m}  \,.
\label{eff-pot}
\end{equation}
Here, ${\cal Q}$ is vanished. The effective potential depends on $E$, $L_z$ and $K$.
For different values of them, ${\cal V}_{\rm effp} (r)$s are different ones.
The local maximum of the potential determines
the radii of unstable circular orbits. For those orbits, ${\cal V}_{\rm effp}=0$ and $\frac{d{\cal V}_{\rm effp}}{dr}=0$,
which is equivalent to find $R(r)=0$ and $dR(r)/dr=0$.

We note that finding homoclinic orbits correspond to finding the unstable circular orbits.
The location of a peak is determined as
\begin{eqnarray}
r_{\rm uco} &=& \Big[3M - \frac{2Q^2}{r_{\rm uco}} + K(1+w)r^{(1-2w)}_{\rm uco}\Big] \nonumber \\
&+&\frac{1}{L^2_z} \Big[(m^2M-eQE)r^2_{\rm uco} + Q^2(e^2-m^2)r_{\rm uco} + Kwm^2 r^{(3-2w)}_{\rm uco} \Big] \,.
\end{eqnarray}
This reduces to the case of the null geodesic motion when $m=e=0$,
while this reduces to the case of the neutral particle motion when $e=0$.
This reduces to the case of Reissner-Nordstr\"om black hole (RN BH) when $K=0$.

For unstable circular orbits, one could obtain
\begin{eqnarray}
E &=& \frac{A(r) \pm B(r)}{2r(2Q^2 + r^2 -3Mr) - 2Kr^{3-2w}(1+w)} \,, \nonumber \\
L_z &=& \frac{r\sqrt{C(r) +D(r)}}{\sqrt{2}[2Q^2+r^2-3Mr-Kr^{2(1-w)}(1+w)]}  \,.
\end{eqnarray}
where
\begin{eqnarray}
A(r) &=& e Q[3Q^2 + r^2 -4Mr -Kr^{2(1-w)}(1+2w)] \,,\nonumber \\
B(r) &=& (Q^2 +r^2 -2Mr -Kr^{2(1-w)}) \sqrt{e^2 Q^2 +4m^2[2Q^2-3Mr +r^2 -K r^{2(1-w)}(1+w)]} \,,\nonumber \\
C(r) &=& e^2Q^2 [Q^2+r^2-2Mr-2Kr^{2(1-w)}(1+w)] \\
     & & -2m^2 (Q^2 -Mr-Kr^{2(1-w)}w)[2Q^2 +r^2 -3Mr -K r^{2(1-w)}(1+w)  ] \,, \nonumber \\
D(r) &=&  -eQ \left[eQK r^{2(1-w)}(1+2w) \right. \nonumber \\
     & & \pm \left.  (Q^2 +r^2 -2Mr -Kr^{2(1-w)})\sqrt{e^2Q^2 +4m^2(2Q^2-3Mr+r^2)-4Km^2r^{2(1-w)}(1+w)} \right]  \nonumber  \,.
\end{eqnarray}

\begin{figure}[H]
\begin{center}
\subfigure[Effective potentials for SBH]
{\includegraphics[scale=0.32]{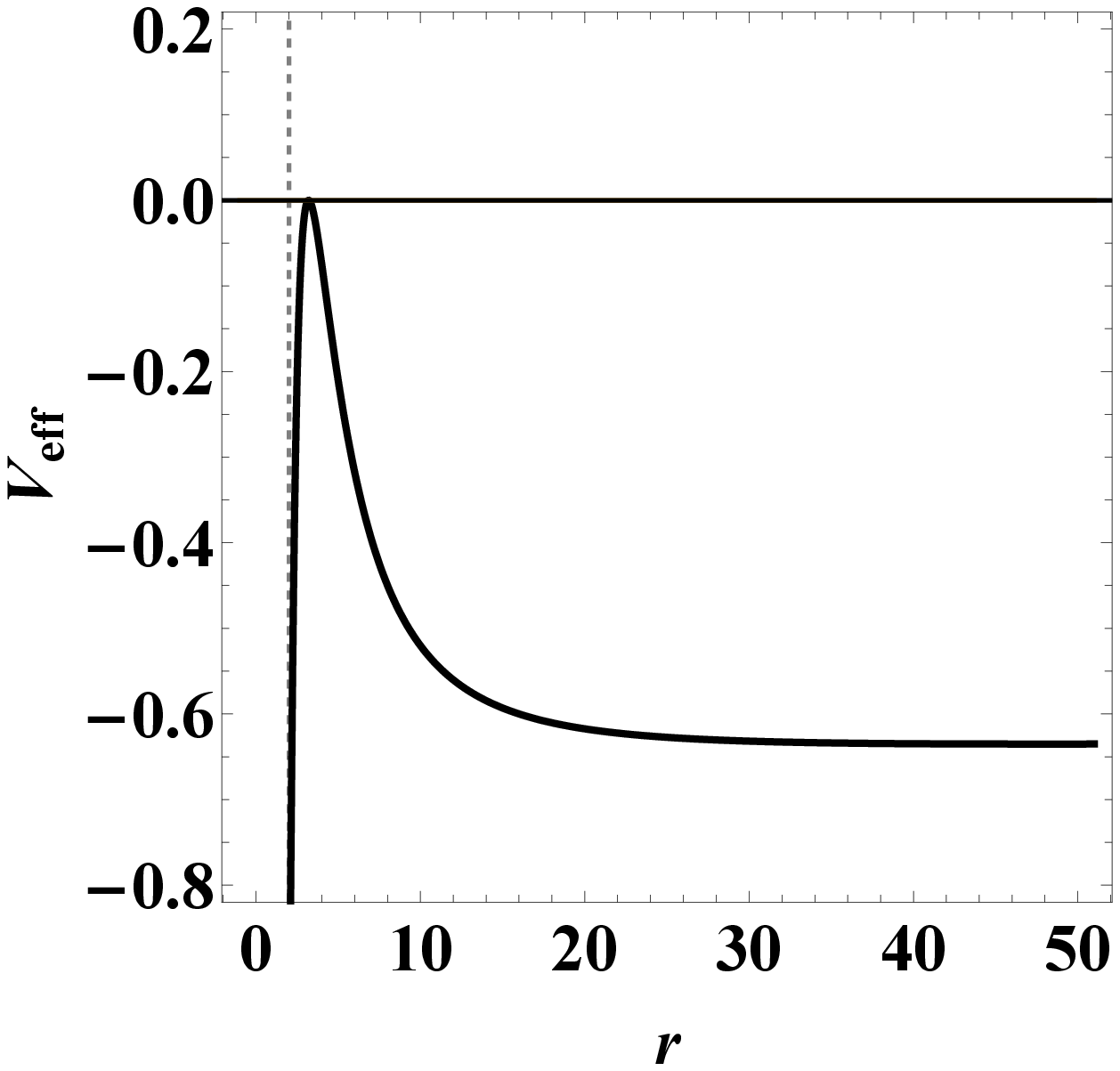}}
\subfigure[Trajectory of the particle in the SBH geometry]
{\includegraphics[scale=0.32]{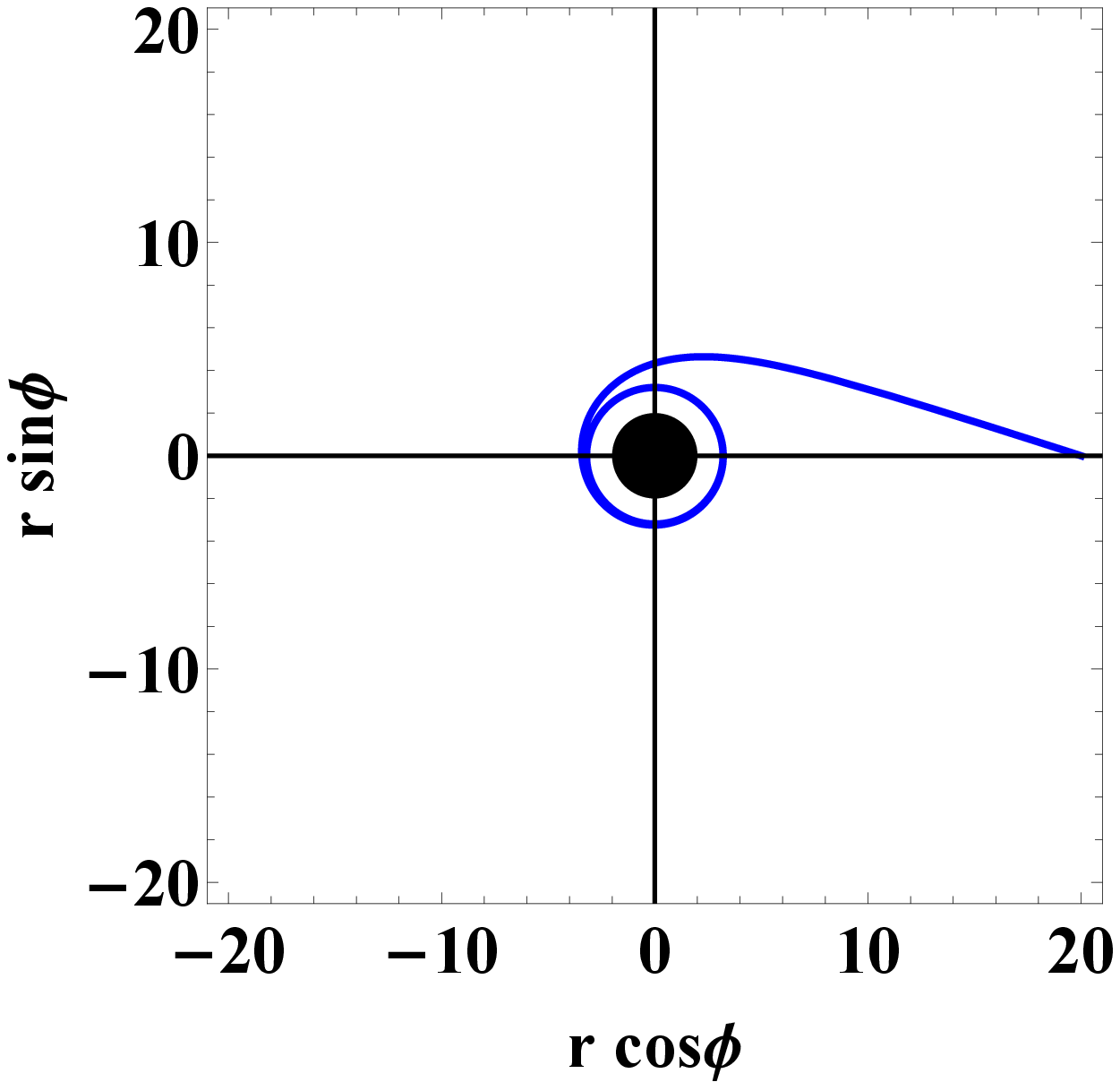}}
\subfigure[Effective potentials for RNBH]
{\includegraphics[scale=0.32]{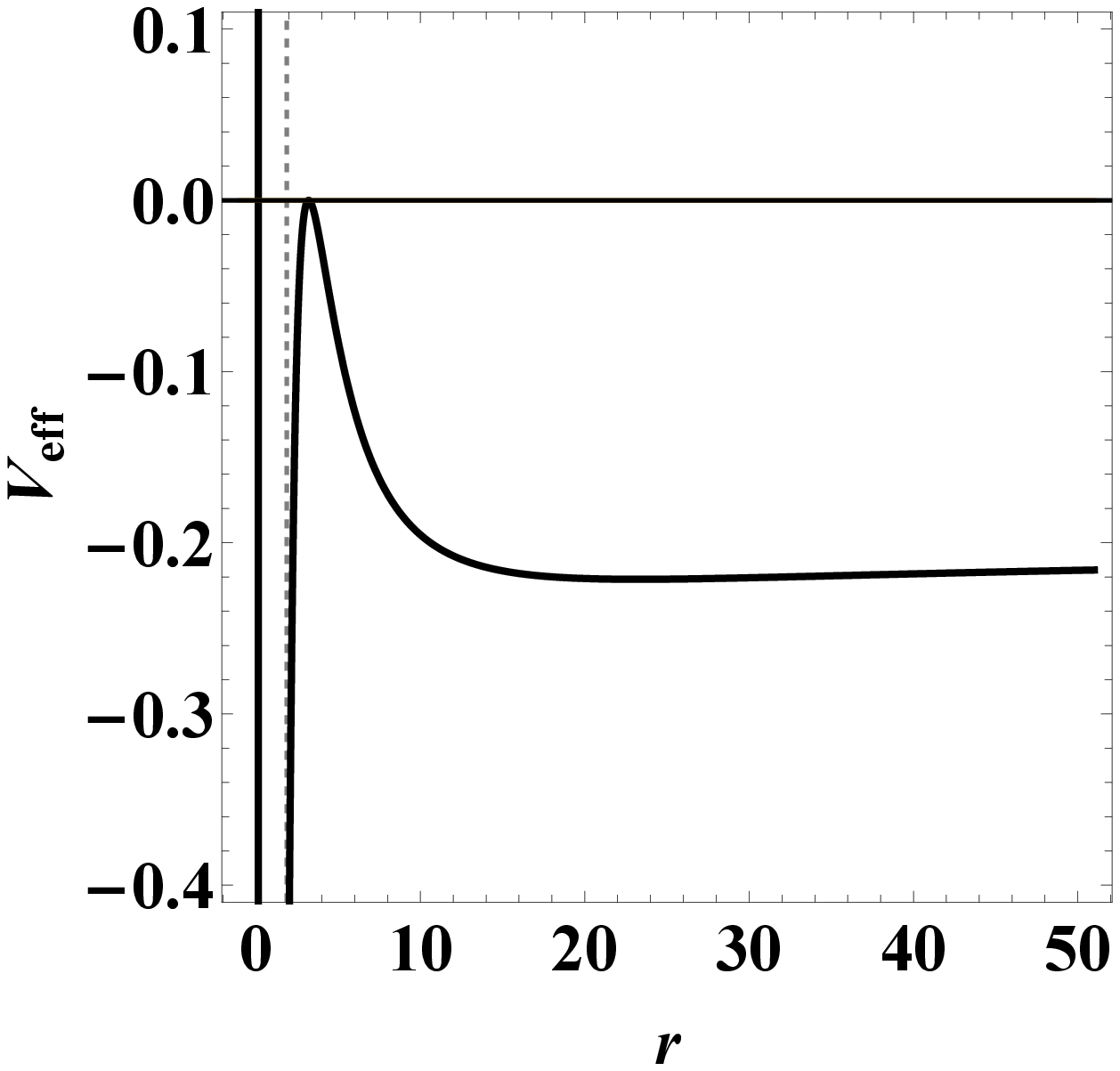}}
\subfigure[Trajectory of the particle in the RNBH geometry]
{\includegraphics[scale=0.32]{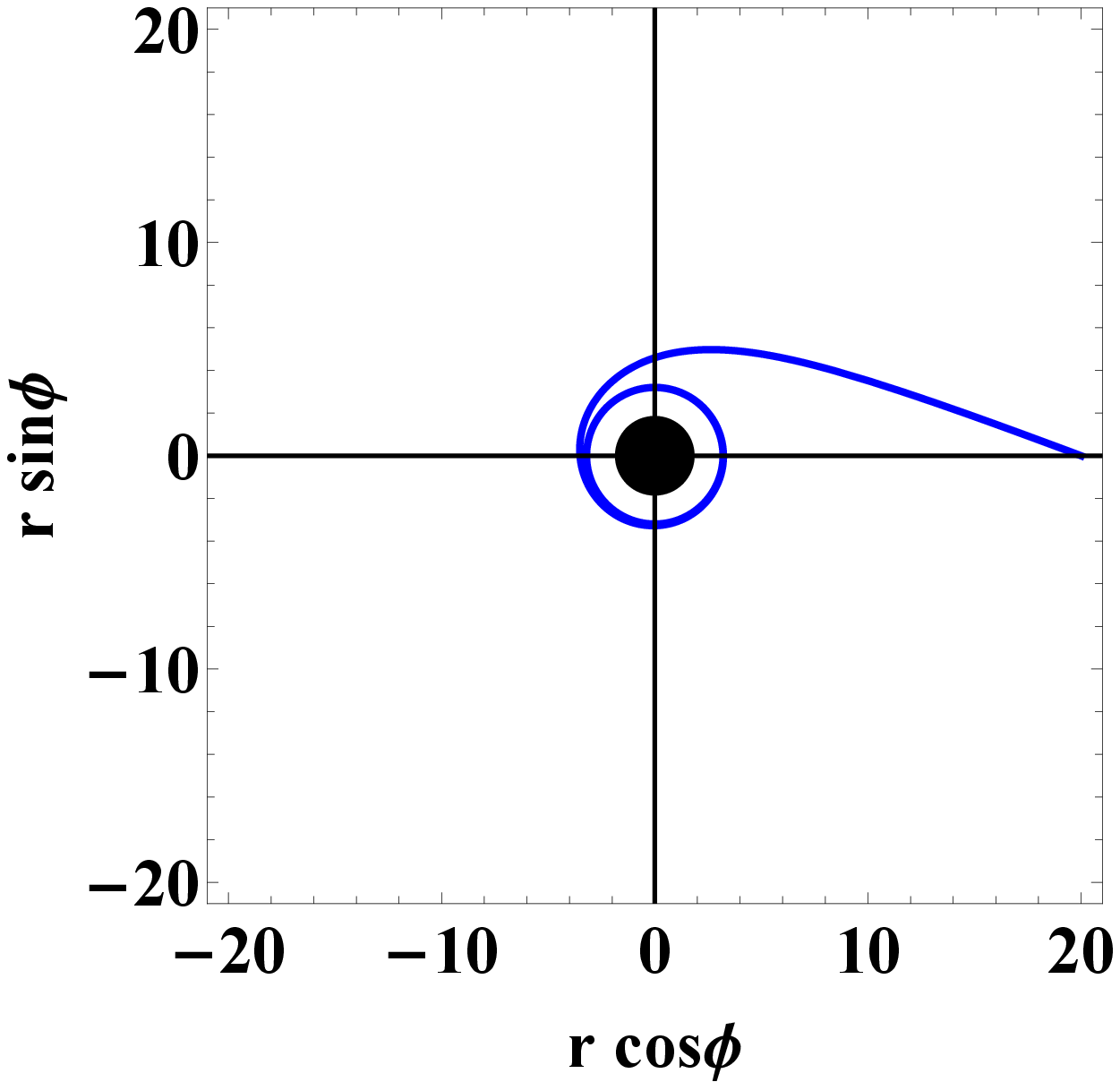}}
\subfigure[Effective potentials for BH with $w=2/3$]
{\includegraphics[scale=0.32]{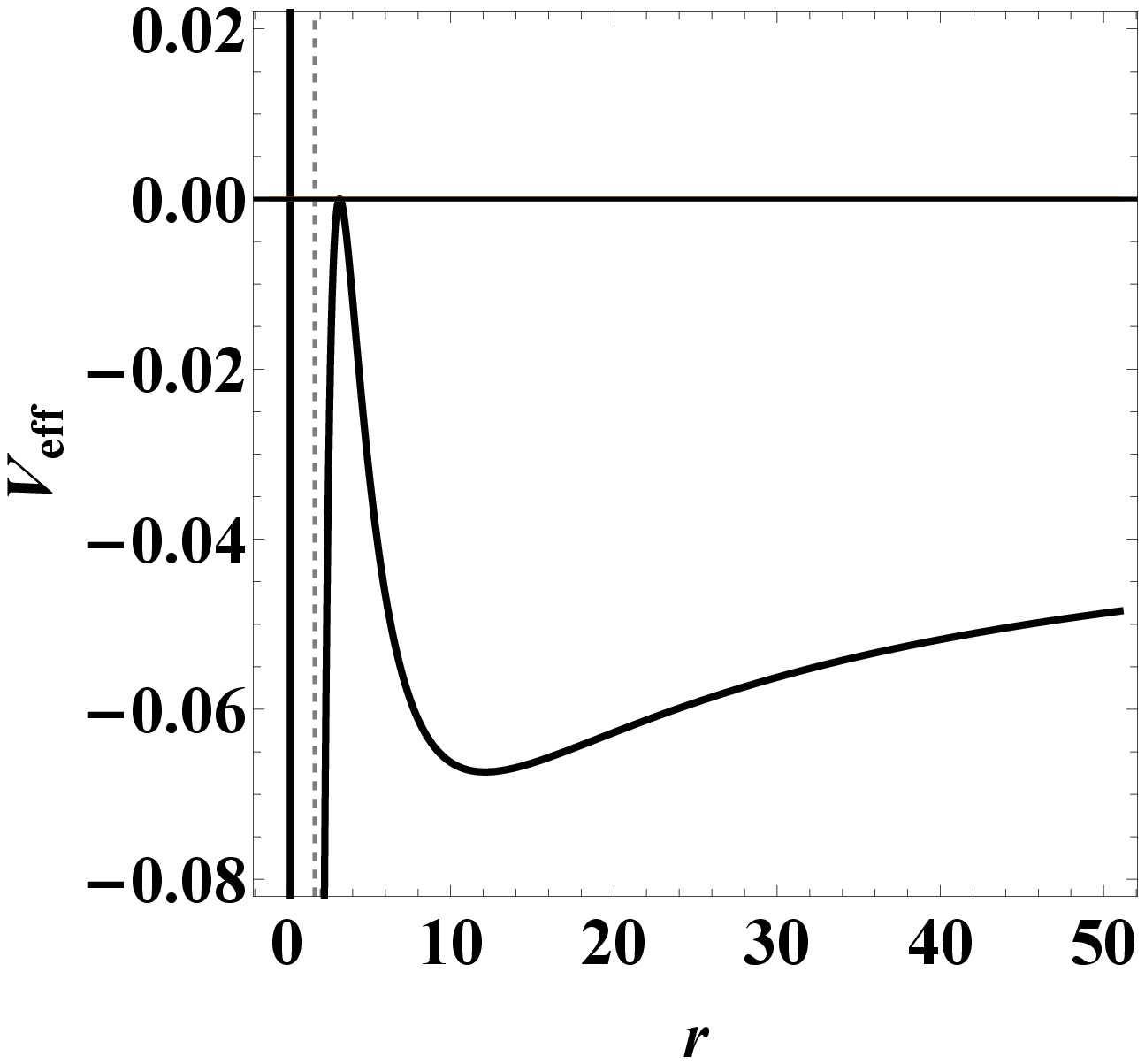}}
\subfigure[Trajectory of the particle in the BH geometry with $w=2/3$]
{\includegraphics[scale=0.32]{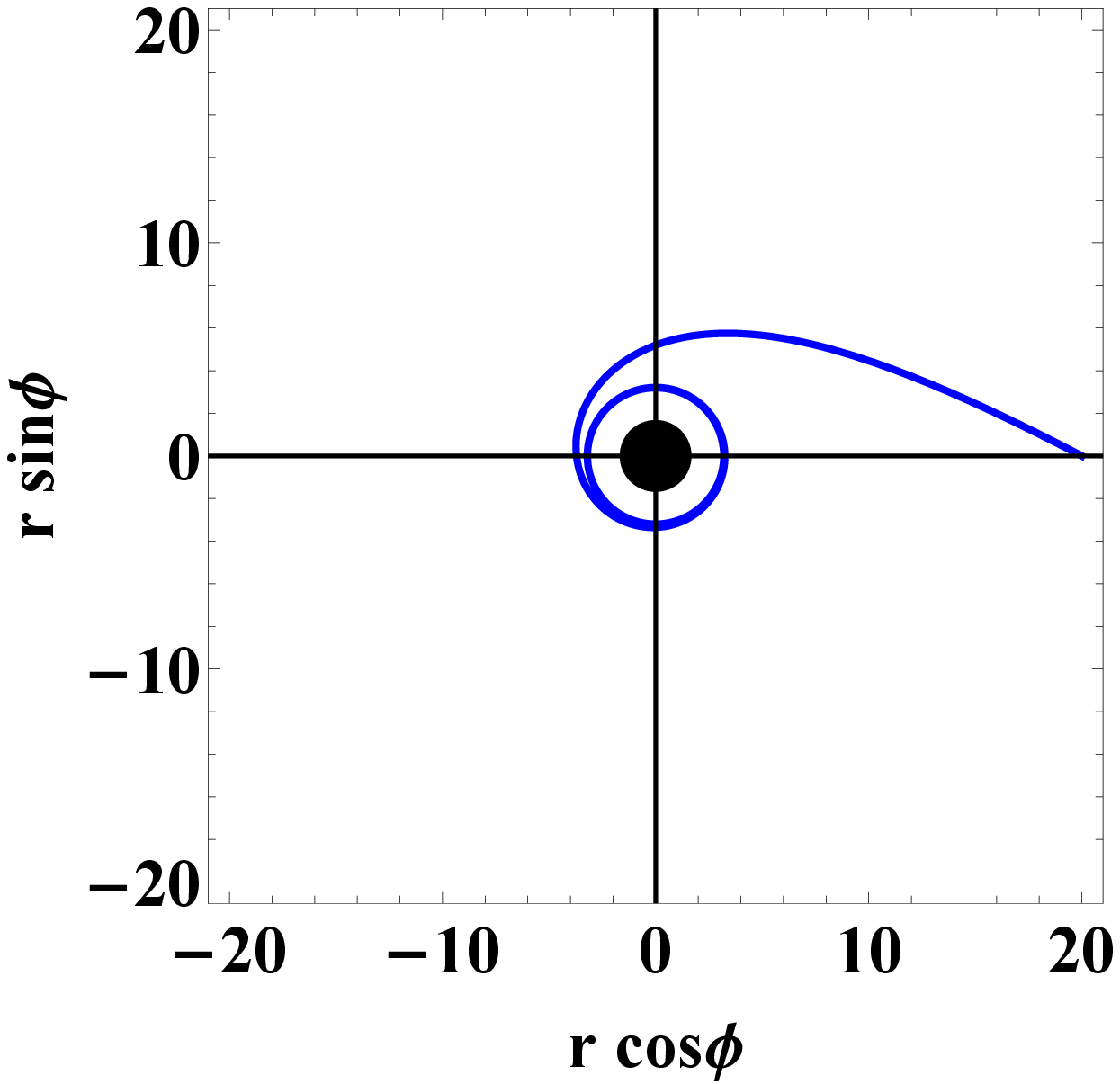}}
\subfigure[Effective potentials for BH with $w=3/2$]
{\includegraphics[scale=0.32]{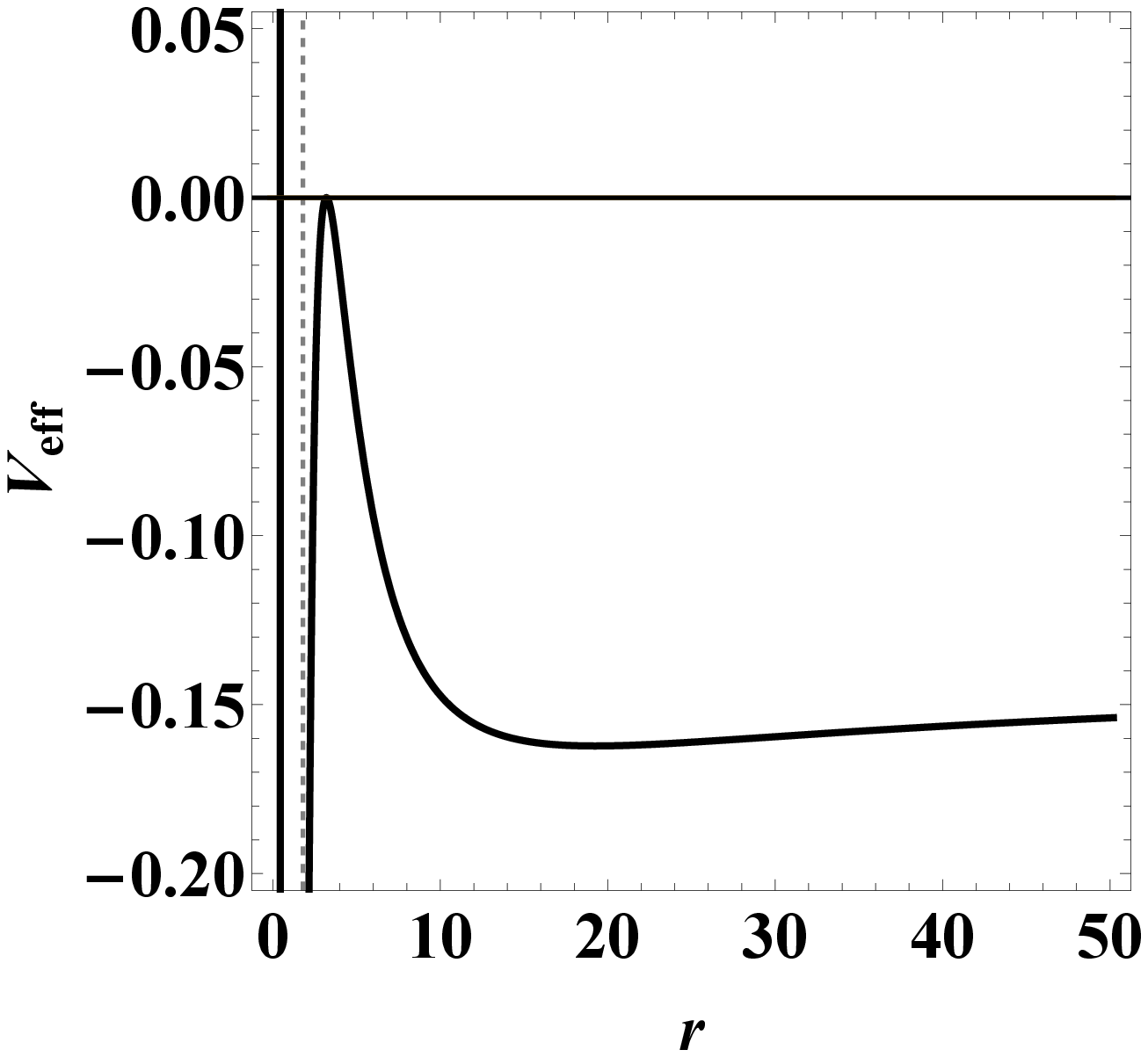}}
\subfigure[Trajectory of the particle in the BH geometry with $w=3/2$]
{\includegraphics[scale=0.32]{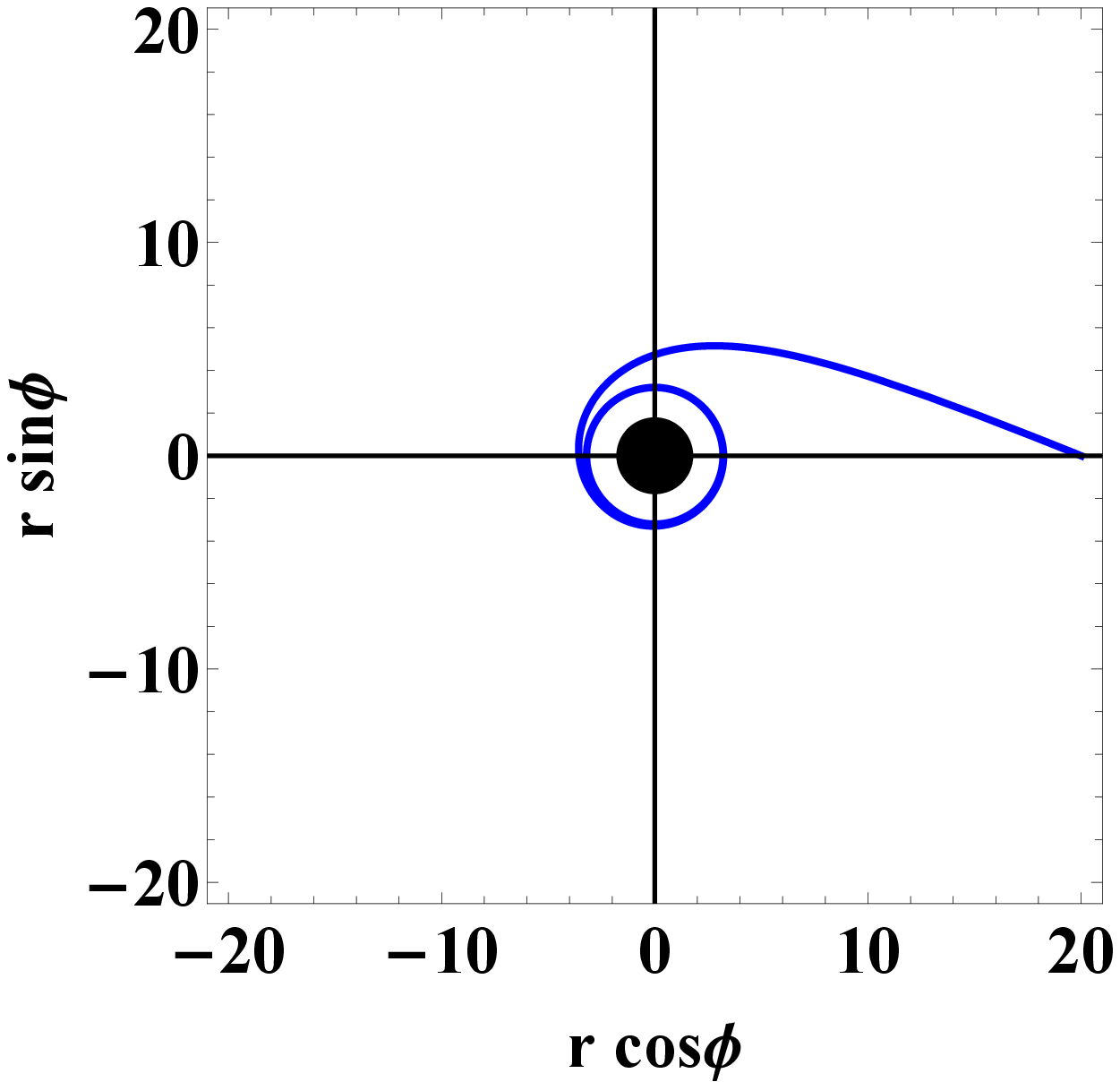}}
\subfigure[Effective potentials for BH with $w=2$]
{\includegraphics[scale=0.32]{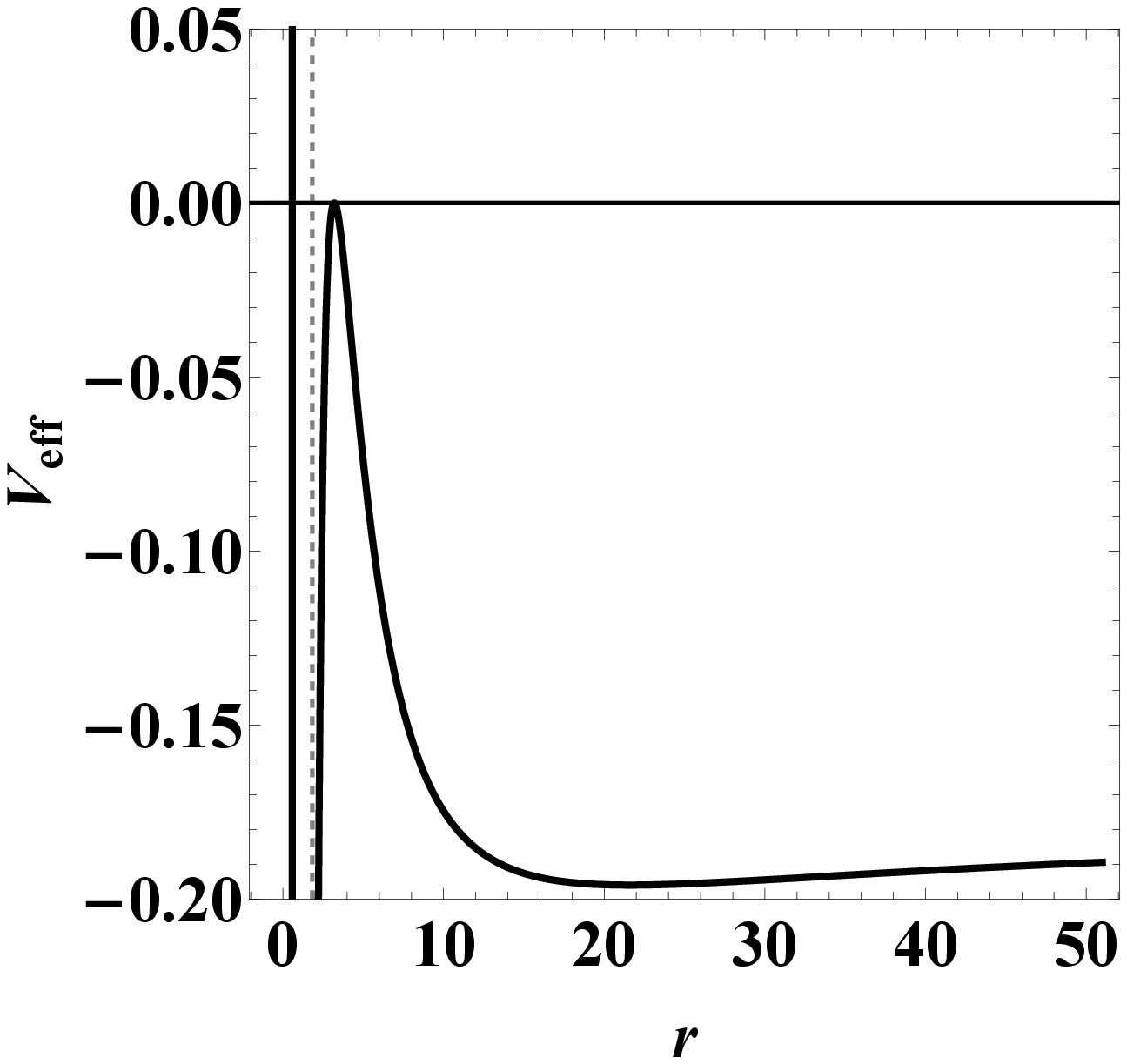}}
\subfigure[Trajectory of the particle in the BH geometry with $w=2$]
{\includegraphics[scale=0.32]{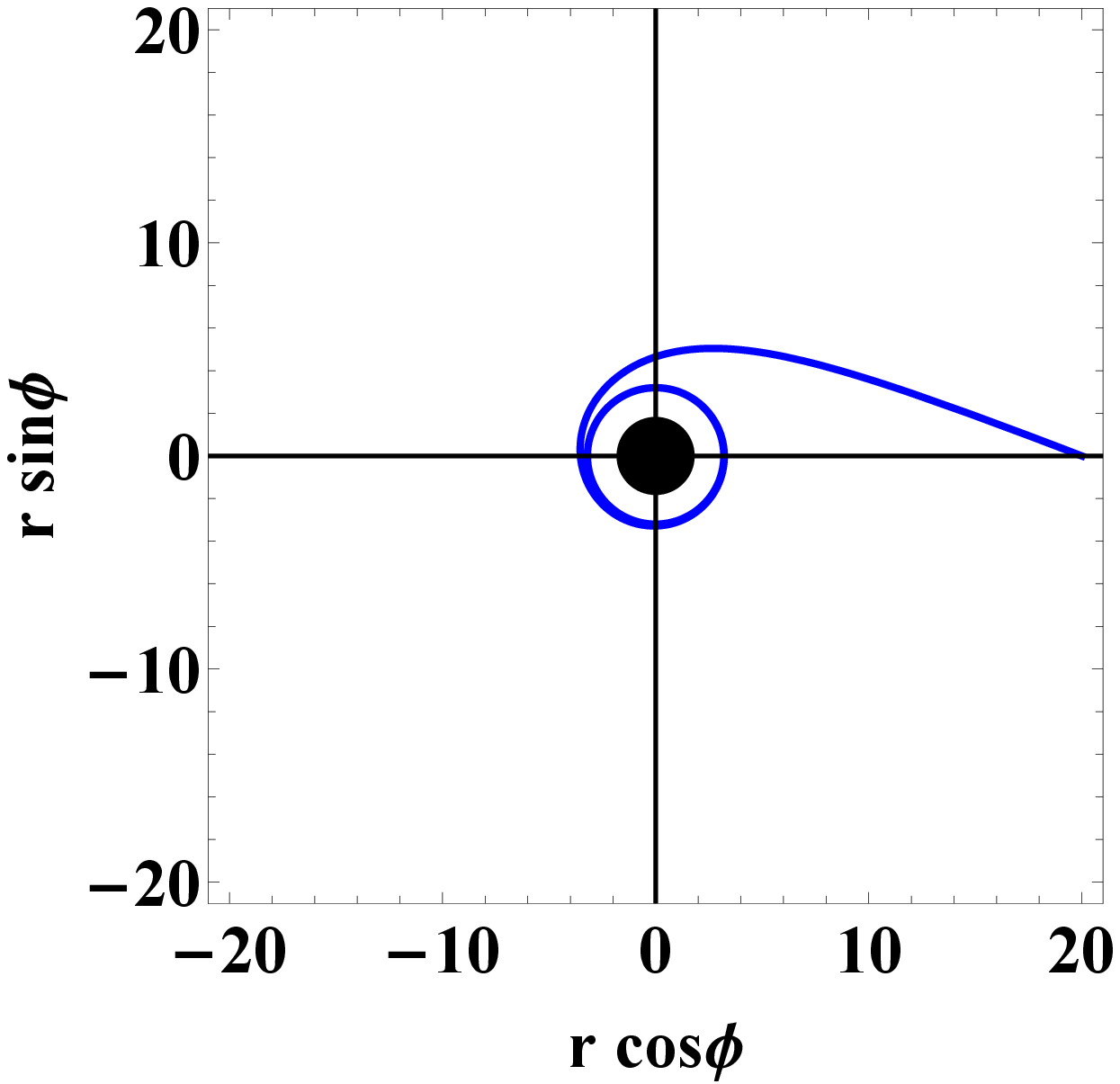}}
\subfigure[Effective potentials for naked RNBH.]
{\includegraphics[scale=0.32]{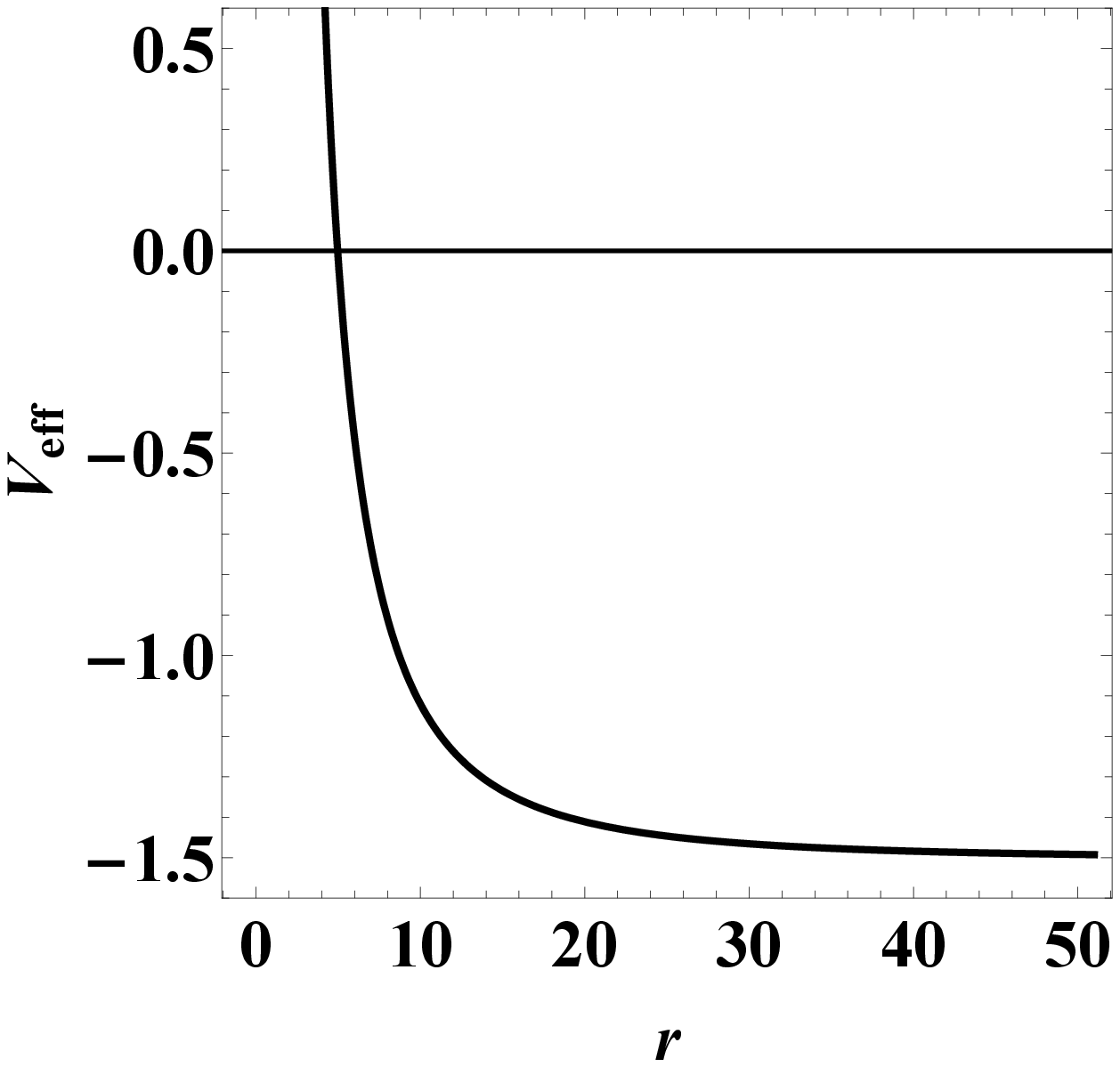}}
\subfigure[Trajectory of the particle in the naked RNBH geometry]
{\includegraphics[scale=0.32]{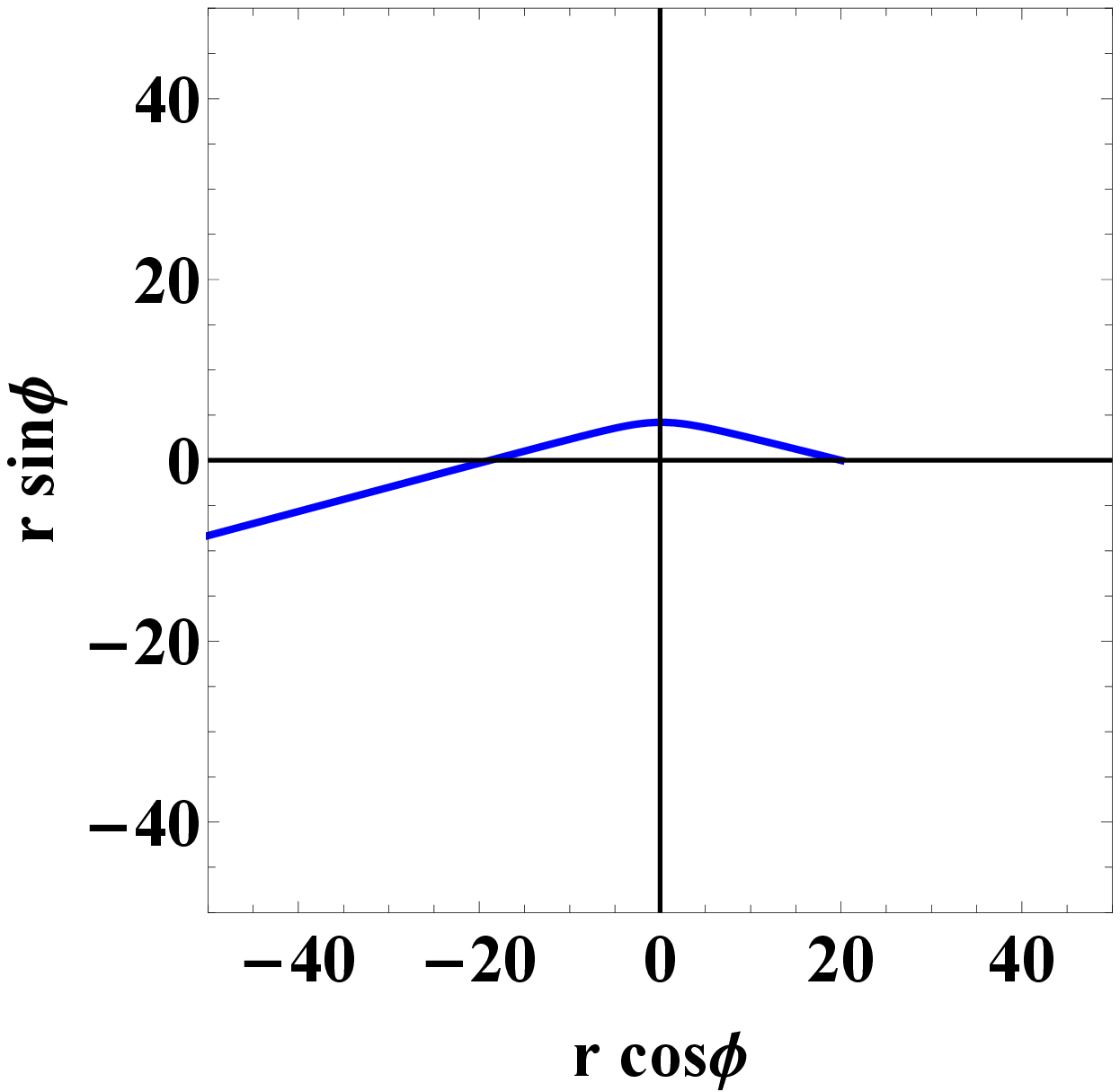}}
\end{center}
\caption{\footnotesize{(color online).
Effective potentials and particle trajectories }}
\label{Effective_potential}
\end{figure}

Figure \ref{Effective_potential} shows the effective potentials,
given by (\ref{eff-pot}), for the cases of homoclinic orbits and a black hole with naked singularity.
In the potential figures, the vertical dotted line indicates the location of the black hole outer horizon
and the local maximum is outside of that.
The location of the local maximum is satisfied to ${\cal V}_{\rm effp} (r)=0$.
To do so, we take the energy value of the particle as different in each case.
Thus, we show each figure separately since it does not make much sense to put all data
together in one figure and compare them. The orbits show those in Schwarzschild black hole (SBH), RNBH with $K=0$,
and BH with anisotropic matter fields with $w=2/3$, $w=3/2$, and $w=2$ in the equatorial plane.
We take $M=1$, $Q=0.5$, $m=1$, $e=0.1$ for all cases.
We take the energy for $w=2/3$ with $E=1.0325$ and $L=3.7044$, $w=3/2$ with $E=0.8544$ and $L=4.5994$,
and $w=2$ with $E=0.8744$ and $L=4.8272$, and $K=-0.2$ for all cases except for SBH and RNBH.
If the energy is higher than we take the value, then the test particle could directly enter the event horizon,
while if that is less, then the particle has different orbits from the homoclinic one.
In figures, we only drew them for a finite time.
However, it takes an infinite time to approach a homoclinic orbit.

A local maximum of the effective potential does not appear in Newtonian mechanics.
On top of that, a black hole with naked singularity needs to be analyzed separately~\cite{Virbhadra:2002ju, Virbhadra:2007kw}.
One could divide that into two types as the charge increases.
$(i)$ Naked singularity will appear, but there exists still the local maximum of the potential.
$(ii)$ Naked singularity exists but there is no local maximum. As a result, the particle could not have the homoclinic orbit.
For instance, $(k)$ and $(l)$ in Fig.~\ref{Effective_potential} show such a case for RNBH.
In the case of $(ii)$, there is no need to investigate the Lyapunov exponent.

\section{Lyapunov exponent and the violation of chaos bound \label{sec3}}

\quad

We now consider the small perturbation of a test particle around the local maximum of the effective potential
described by the black hole metric (\ref{metricfunction}).
The slightly perturbed orbit is described by $r(\lambda)=r_o+\varepsilon(\lambda)$.
For the radial perturbation, one could obtain the position of that extremum $r_o$ by solving ${\cal V}'_{\rm effp}(r)|_{r=r_o}=0$,
in which the prime denotes the derivatives with respect to $r$.
We are interested in the behavior of particle orbits near the black hole.

There is an issue with choosing the time~\cite{Wu:2003pe, Perez-Giz:2008ajn}, resulting in the changing in
the potential shape due to choosing that coordinate.
We choose the proper time $\tau$, when we consider one geodesic orbit as the homoclinic orbit.
When we consider the orbits of two particles, we take the coordinate time $t$ rather
than the proper time $\tau$ as the independent variable,
since the two particles have different proper times.
We note that the coordinate time has to be adopted since one independent variable
should be adopted when their equations of motion are integrated numerically.
As a result, the shape of the potential could be changed slightly.

We now investigate the Lyapunov exponent as a strong indicator showing the violation of chaos bound.
This one measures the sensitivity to which orbits move away with time.
If its value is positive, it is worthwhile to investigate the possibility of the violation of chaos bound.

We consider two initial points $r_o$ and $r_o + \varepsilon$.
The Lyapunov exponent $\lambda$ corresponds to the measure showing
the average exponential growth per unit time between two nearby trajectories,
reflecting a high sensitivity to initial conditions.
The difference between two trajectories is
approximately given by
\begin{equation}
d(t) \sim \varepsilon e^{\lambda t} \,. \label{caosbehavior}
\end{equation}
We note that if $\lambda$ is negative, then the two orbits will eventually converge,
while if positive, they diverge.

In a thermal quantum system, the exponent $\lambda$
is bounded as~\cite{Maldacena:2015waa}
\begin{equation}
\lambda \leq \frac{2\pi T}{\hbar}  \,, \nonumber
\end{equation}
and it was applied to a black hole system~\cite{Hashimoto:2016dfz}
\begin{equation}
\lambda \leq \frac{2\pi T_H}{\hbar} =\kappa \,, \label{Lyapunov}
\end{equation}
where $T_H$ denotes the black hole temperature and $\kappa$ is the surface gravity
of the black hole. The Eq.\ (\ref{Lyapunov}) is equivalent as follows:
\begin{equation}
\kappa^2 - \lambda^2 \geq 0 \,. \label{Lyapunov2}
\end{equation}

From Eqs.\ (\ref{HJeq-t}) and (\ref{HJeq-the}), we obtain the radial equation in terms of the coordinate time,
\begin{equation}
\frac{1}{2} m \left(\frac{dr}{d t} \right)^2  + {\cal V}_{\rm effc} (r)= 0 \,, \label{eqgeocot}
\end{equation}
where the effective potential in the equatorial plane is given by
\begin{equation}
{\cal V}_{\rm effc} (r)= \frac{f^2(r)[f(r)[m^2 + \frac{L^2_z}{r^2}]-(E-\frac{eQ}{r})^2]}{2m(E-\frac{eQ}{r})^2 }  \,.
\label{eff-pot-cot}
\end{equation}
The Eq.\ (\ref{eqgeocot}) becomes
\begin{eqnarray}
 0 & \simeq&  \frac{1}{2} m \left(\frac{d\varepsilon}{dt}\right)^2 +\left( {\cal V}_{\rm effc}(r_o)
+ \frac{1}{2}{\cal V}''_{\rm effc}(r_o) \varepsilon^2 \right)
+ {\cal O}(\epsilon^3)   \nonumber \\
 &=& \frac{1}{2} m \left[\left(\frac{d \varepsilon}{dt}\right)^2 - \lambda^2 \varepsilon^2\right] \,,
\end{eqnarray}
where we neglected the constant term and higher order terms. The coefficient of $\varepsilon^2$ corresponds to
the Lyapunov exponent as
\begin{equation}
\lambda^2  = - \frac{{\cal V}''_{\rm effc}(r)|_{r=r_o}}{m} \,,
\end{equation}
where
\begin{eqnarray}
{\cal V}''_{\rm effc}(r) &=& \frac{2 f(r)^3 \left(e^2 m^2 Q^2+2 e E m^2 Q r+3 E^2 L^2_z\right)}{2 m (e Q-E r)^4} \nonumber \\
&& +\frac{3 f(r)^2 \left(f''(r) \left(L^2_z+m^2 r^2\right) (e Q-E r)+4 f'(r) \left(e m^2 Q r+E L^2_z\right)\right)}{2 m (e Q-E r)^3} \nonumber \\
&& +\frac{6 f(r) f'(r)^2 \left(L^2_z+m^2 r^2\right)}{2 m (e Q-E r)^2}-2 f'(r)^2-2 f(r) f''(r) \,,
\end{eqnarray}
and
\begin{eqnarray}
{\cal V}''_{\rm effc}(r_o) &=& \frac{f\left(r_o\right){}^3 \left(e^2 m^2 Q^2+2 e E m^2 Q r_o+3 E^2 L^2_z\right)}{m \left(e Q-E r_o\right)^4} \nonumber
\\
&& +\frac{3 f\left(r_o\right)^2 f''\left(r_o\right) \left(L^2_z+m^2 r_o^2\right)}{2 m \left(e Q-E r_o\right)^2}-2 f\left(r_o\right) f''\left(r_o\right) \,,
\end{eqnarray}
where $f'(r)=\frac{Mr -Q^2 + Kw r^{2-2w}}{r^3}$, $f''(r)=\frac{2(-2Mr +3Q^2 - Kw(1+2w)r^{2-2w})}{r^4}$.
If ${\cal V}''_{\rm effc}(r) < 0$, which implies $\lambda^2 > 0$, the extremum corresponds to the local maximum.
Consequentially, the shape of potential becomes similar to that of the inverse harmonic oscillator describing a chaotic system.
Here we assumed the perturbed system to be a Hamiltonian one.

We now examine specific cases with different values of parameters.

{\bf A. The extremal case }

For the extremal case, there exist two horizons coalesce and hence $\kappa=0$. The chaos bound (\ref{Lyapunov}) becomes
\begin{equation}
\kappa^2 - \lambda^2  = - \lambda^2 \geq 0 \,.
\end{equation}
If there exists a local maximum in the effective potential, $\lambda^2 > 0$, then
the bound is violated.

We examine the bound on the Lyapunov exponent
in the near-horizon region~~\cite{Hashimoto:2016dfz, Zhao:2018wkl, Kan:2021blg}.
We introduce the small parameter $\epsilon$ as $r_o=r_+ + \epsilon$ for the study of the extremal black hole.
We take the local maximum of the effective potential in the vicinity of the horizon
since the Lyapunov exponent is zero if the local maximum is at the horizon.
Then, the metric function becomes $f(r)=0+f'(r_+)\epsilon+\frac{1}{2}f''(r_+) \epsilon^2+ \dotsm$.
This gives
\begin{equation}
\lambda^2  = \frac{f'(r_+)^2}{m^2} + \mathcal{O}(\epsilon) = \frac{4 \left( M r - Q^2 + w K r^{2-2w} \right)}{m^2 r^6_+} + \mathcal{O}(\epsilon) \,.
\end{equation}

{\bf B. The nonextremal case }

The square of the bound is given by
\begin{equation}
\lambda^2 \leq \kappa^2 = \frac{1}{4} f'^2(r)|_{r=r_+} = \left[ \frac{w}{r_+} + \frac{M(1-2w)}{r^2_+}
- \frac{Q^2 (1-w)}{r^3_+} \right]^2 \,.
\end{equation}
We consider the case in which the local maximum is in the vicinity of the horizon. The chaos bound becomes
\begin{eqnarray}
\kappa^2 - \lambda^2 = \frac{1}{4} f'^2(r)|_{r=r_+} + \frac{{\cal V}''_{\rm effc}(r)|_{r=r_o}}{m} \,.
\end{eqnarray}

We perform the numerical calculations for $\lambda^2$ and $\kappa^2-\lambda^2$, separately.
First, we analyze $\lambda^2$ in terms of angular momentum and $K$ at a given energy value.
\begin{figure}[H]
\begin{center}
\subfigure[$w=2/3$]
{\includegraphics[width=2. in]{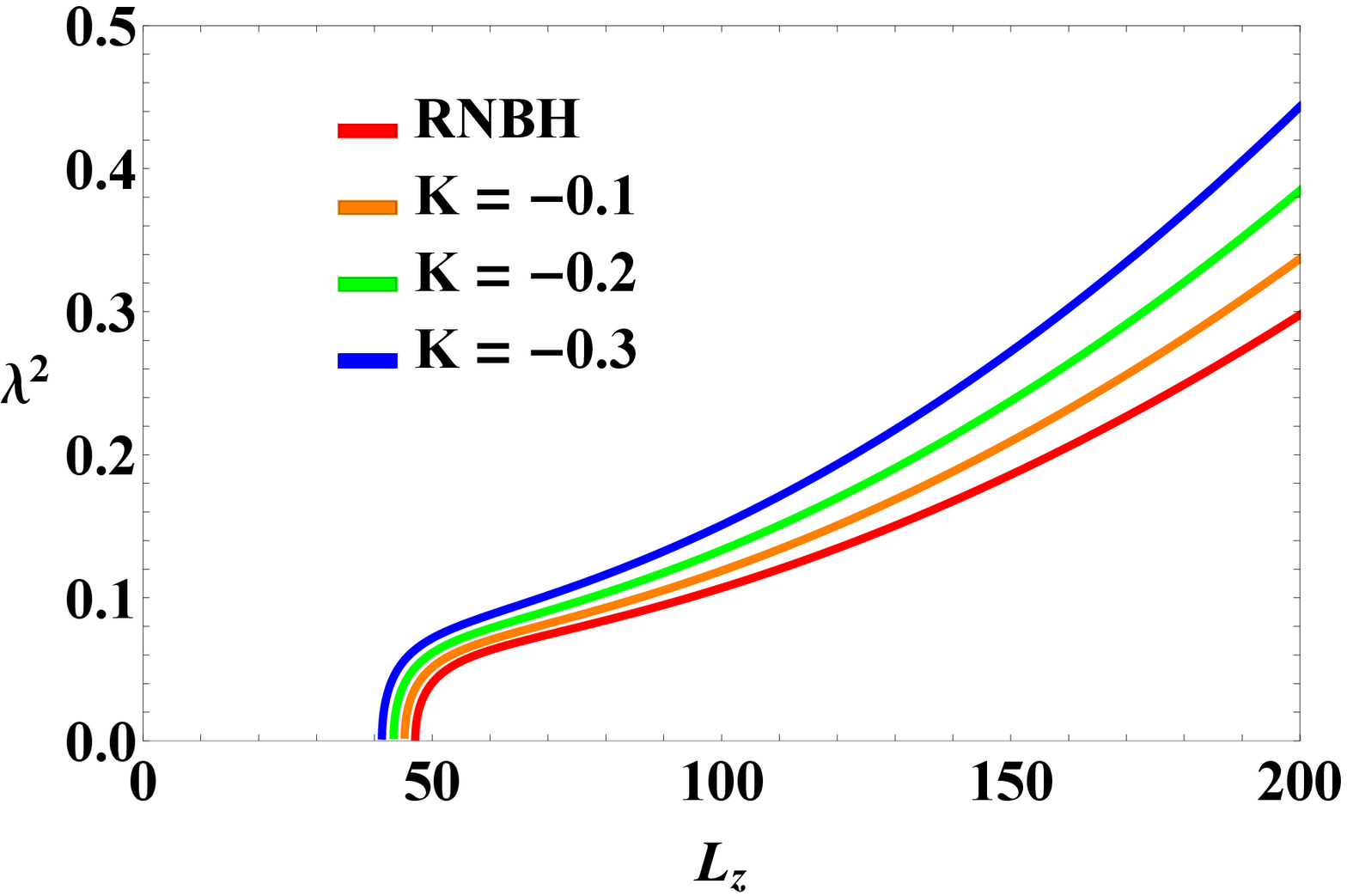}}
\subfigure[$w=3/2$]
{\includegraphics[width=2. in]{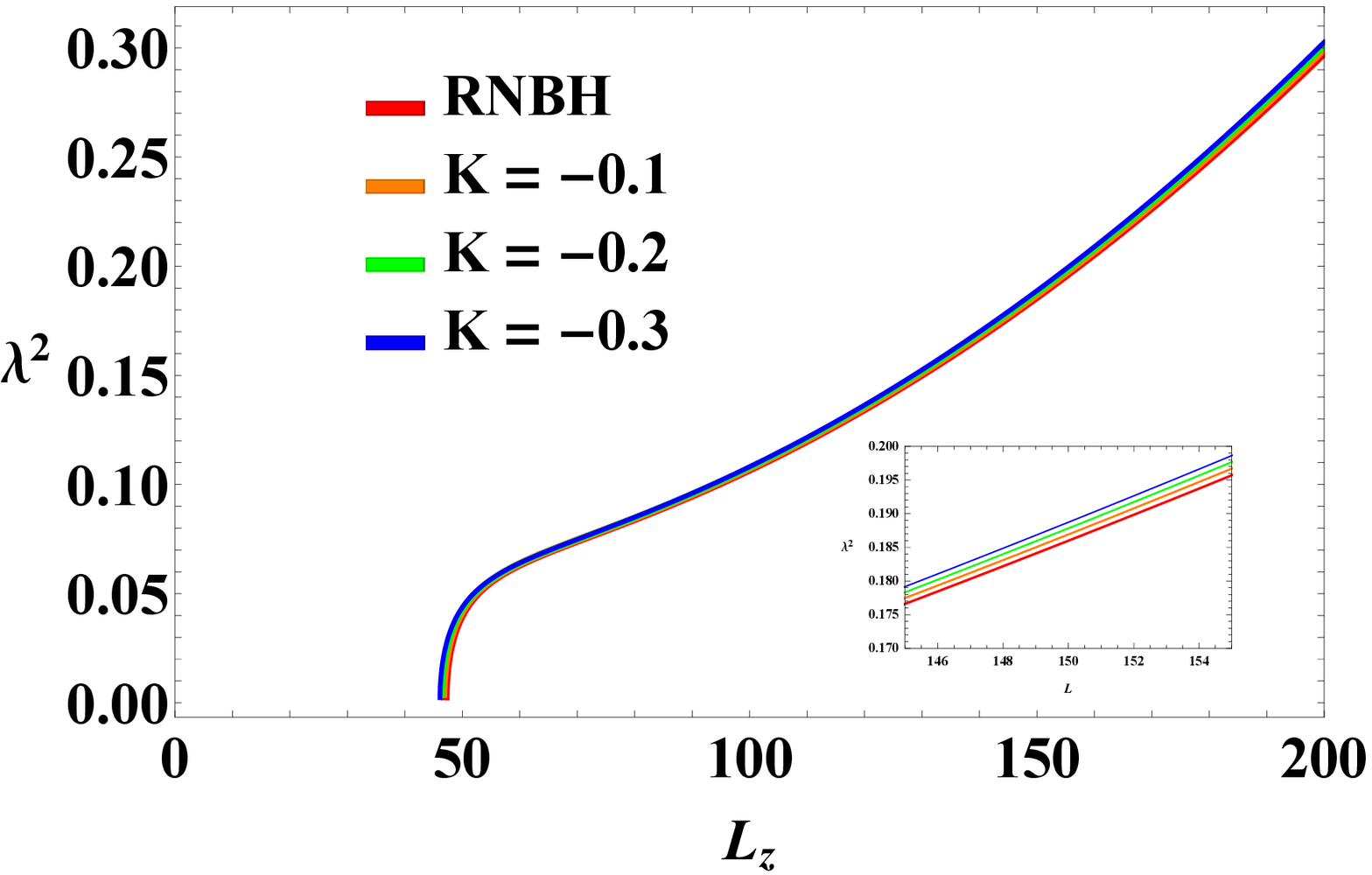}}
\subfigure[$w=2$]
{\includegraphics[width=2. in]{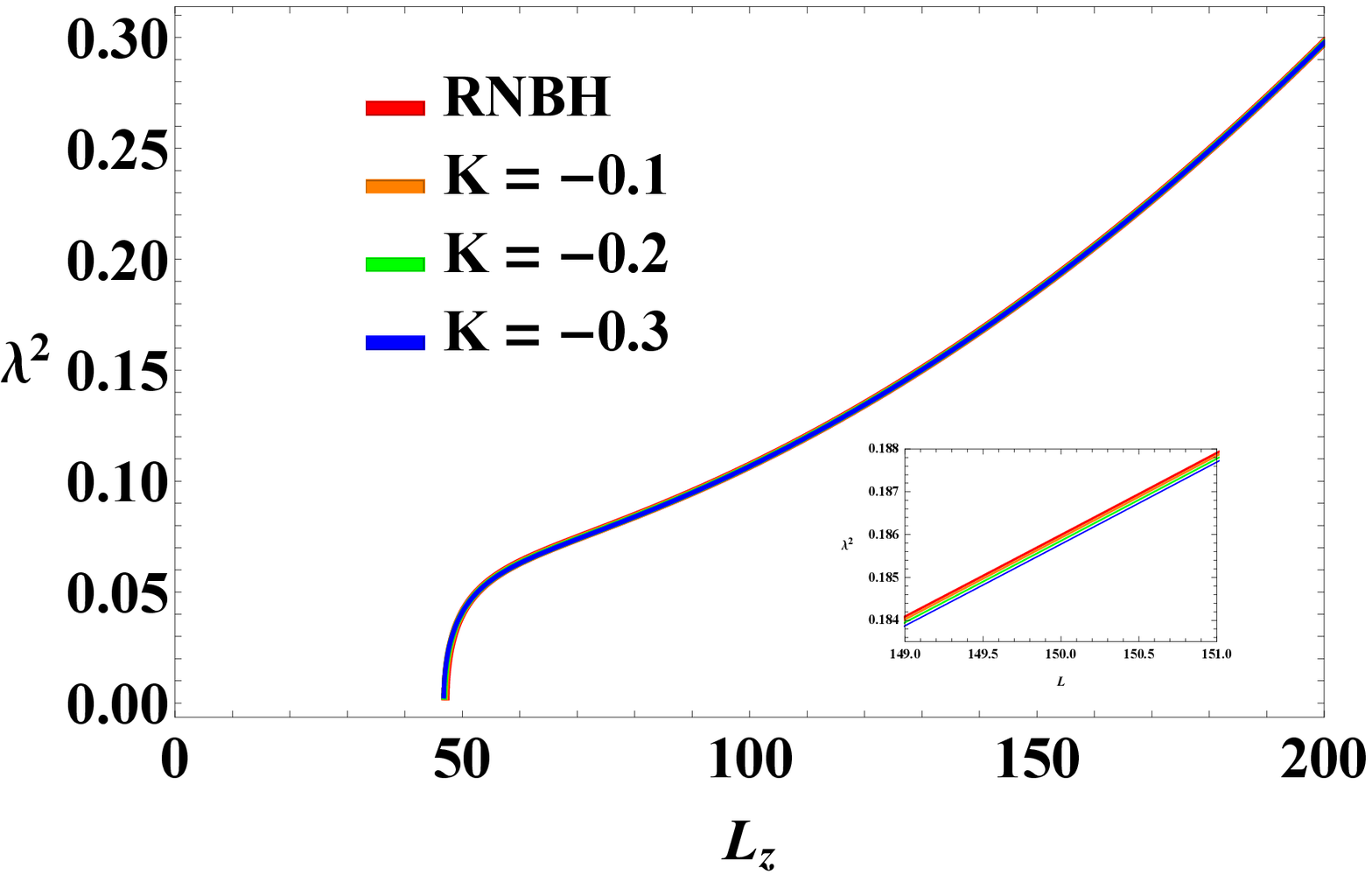}}
\end{center}
\caption{\footnotesize{(color online).
Lyapunov exponent in terms of $L_z$ with different $w$. }}
\label{Lyapunov30}
\end{figure}
Figure \ref{Lyapunov30} shows Lyapunov exponents $\lambda^2$ in terms of $L_z$.
We take $M=1$, $Q=0.5$, $m=1$, $e=0.1$, and $E=10$.
The red line indicates the case of RN ($K=0$), the orange line indicates the case with $K=-0.1$, the green line the case with $K=-0.2$,
and the blue line indicates the case with $K=-0.3$.
Their behaviors show that the value of $\lambda^2$ increases when $L$ or $|K|$ increases.
That is to say, the additional anisotropic matter field enhances the Lyapunov exponent for $w=2/3$ and $3/2$.
For the case with $w=2$, the behavior is different from those with $w=2/3$ and $3/2$.
Where the angular momentum is large, the effect of $K$ seems to appear to be the opposite.
This figure shows that below a specific angular momentum at a given energy, the square of the Lyapunov exponent
does not have a positive value, and at values greater than that value,
when the angular momentum increases, the Lyapunov exponent increases.
The left region below the specific angular momentum corresponds to the region that does not satisfy ${\cal V}'_{\rm effc}=0$
or ${\cal V}''_{\rm effc}< 0$.

In the below, we will show $\lambda^2$ eventually exceeds the $\kappa^2$ and the violation of the bound, $\kappa^2 - \lambda^2 < 0$, occurs.
\begin{figure}[H]
\begin{center}
\subfigure[$w=2/3$ and $K = 0.6 K_{\rm{Max1}}$]
{\includegraphics[width=2. in]{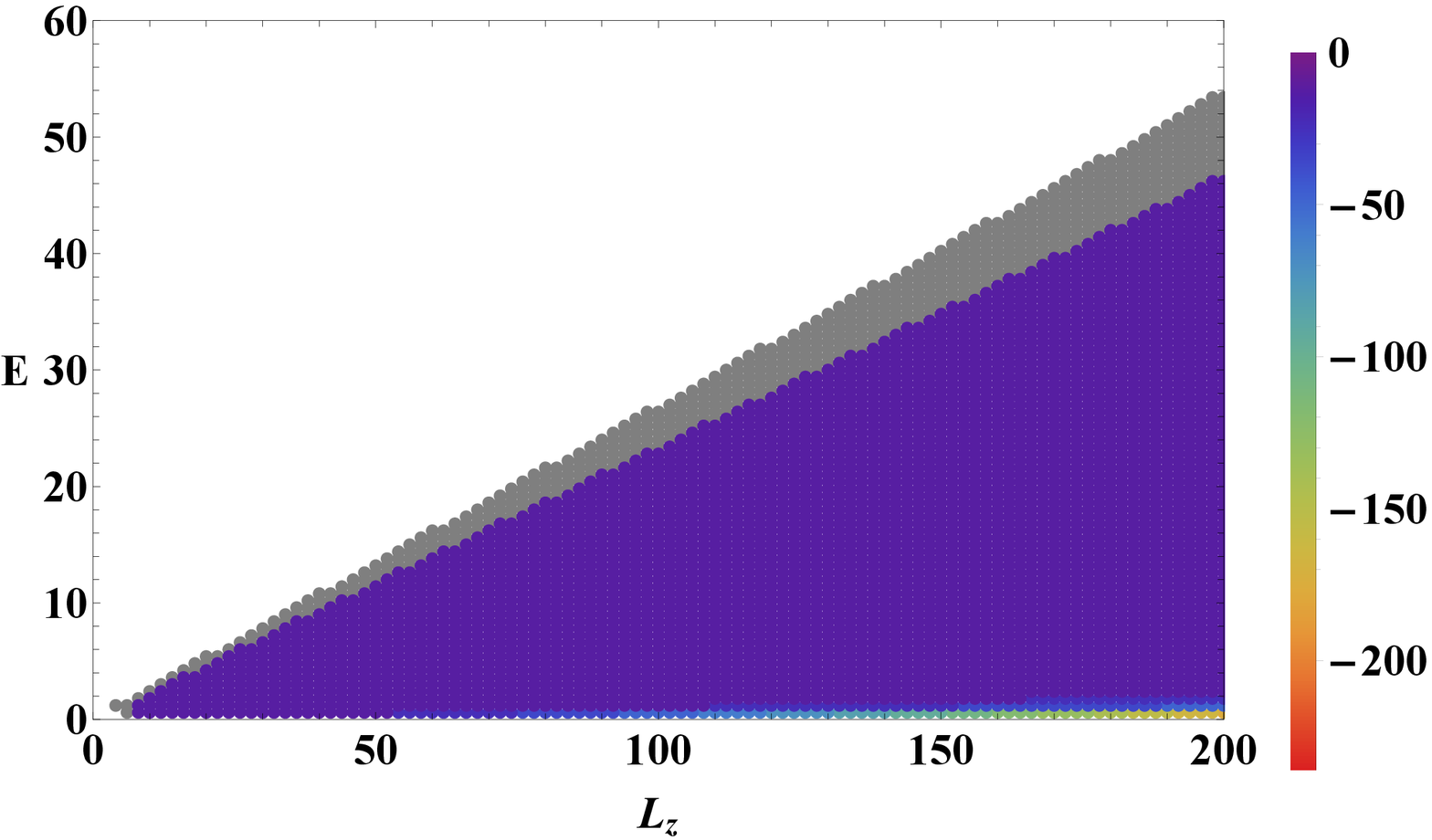}}
\subfigure[$w=2/3$ and $K = 0.8 K_{\rm{Max1}}$]
{\includegraphics[width=2. in]{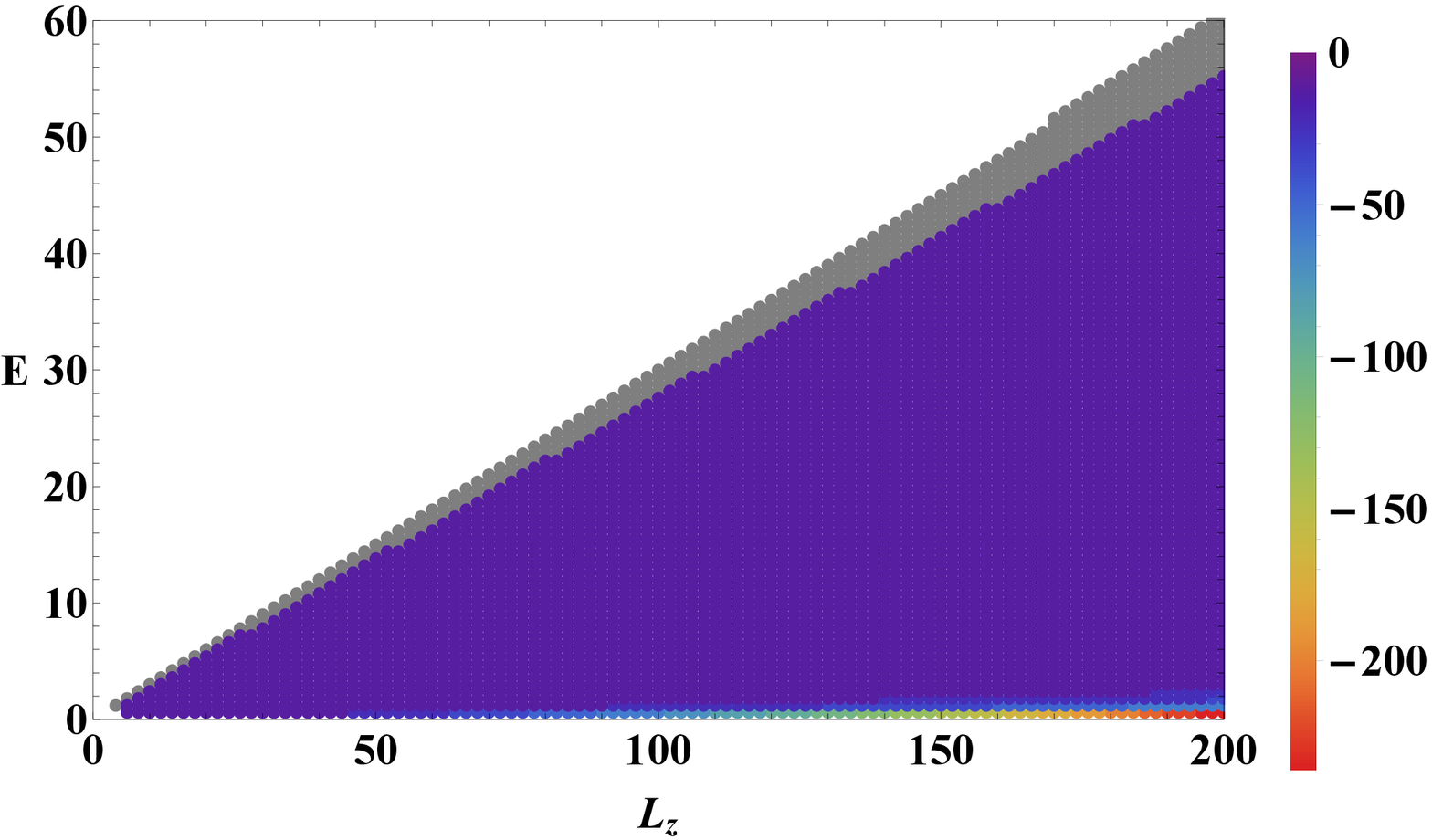}}
\subfigure[$w=2/3$ and $K = K_{\rm{Max1}}$]
{\includegraphics[width=2. in]{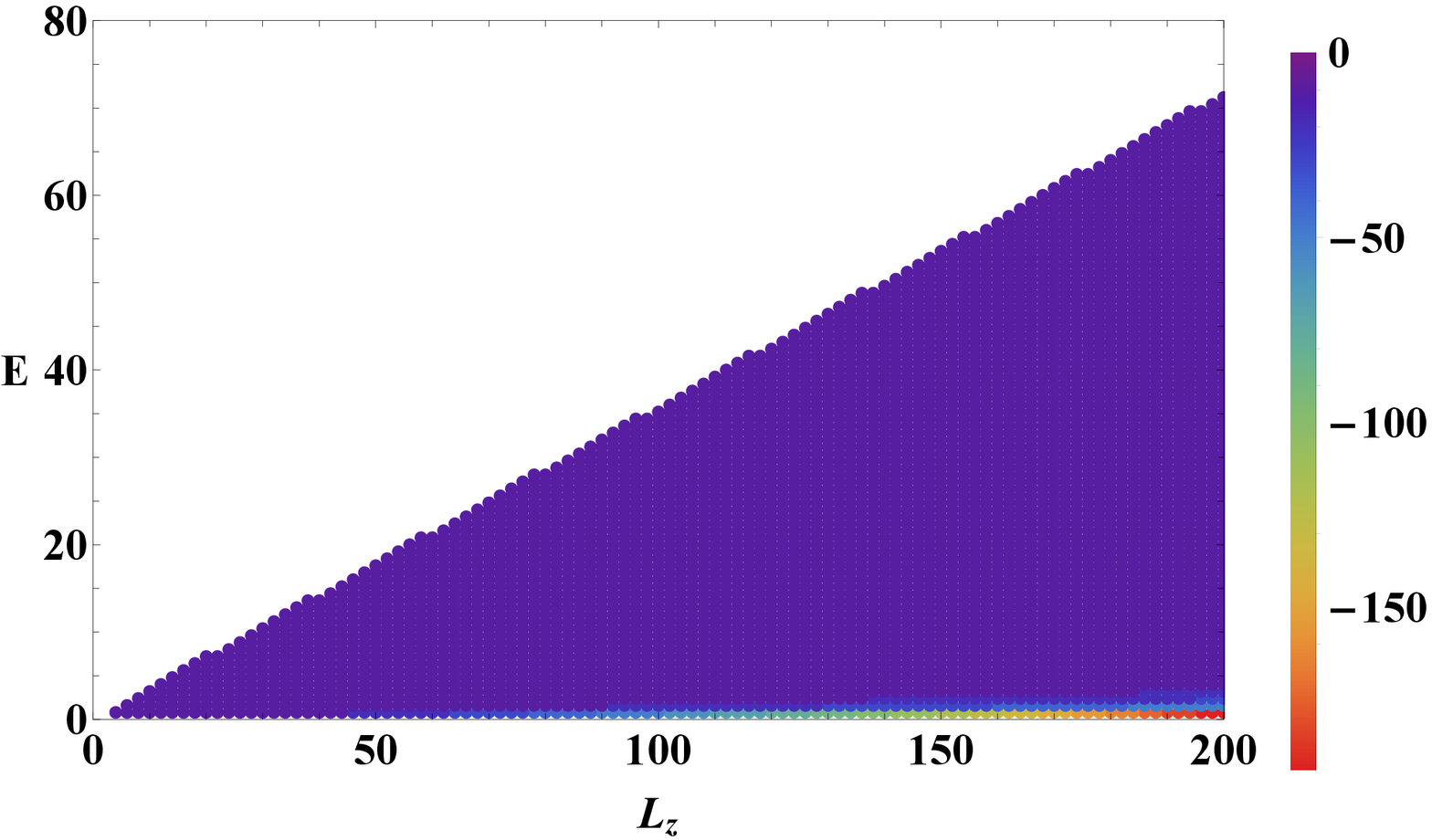}}
\subfigure[$w=3/2$ and $K = 0.6 K_{\rm{Max2}}$]
{\includegraphics[width=2. in]{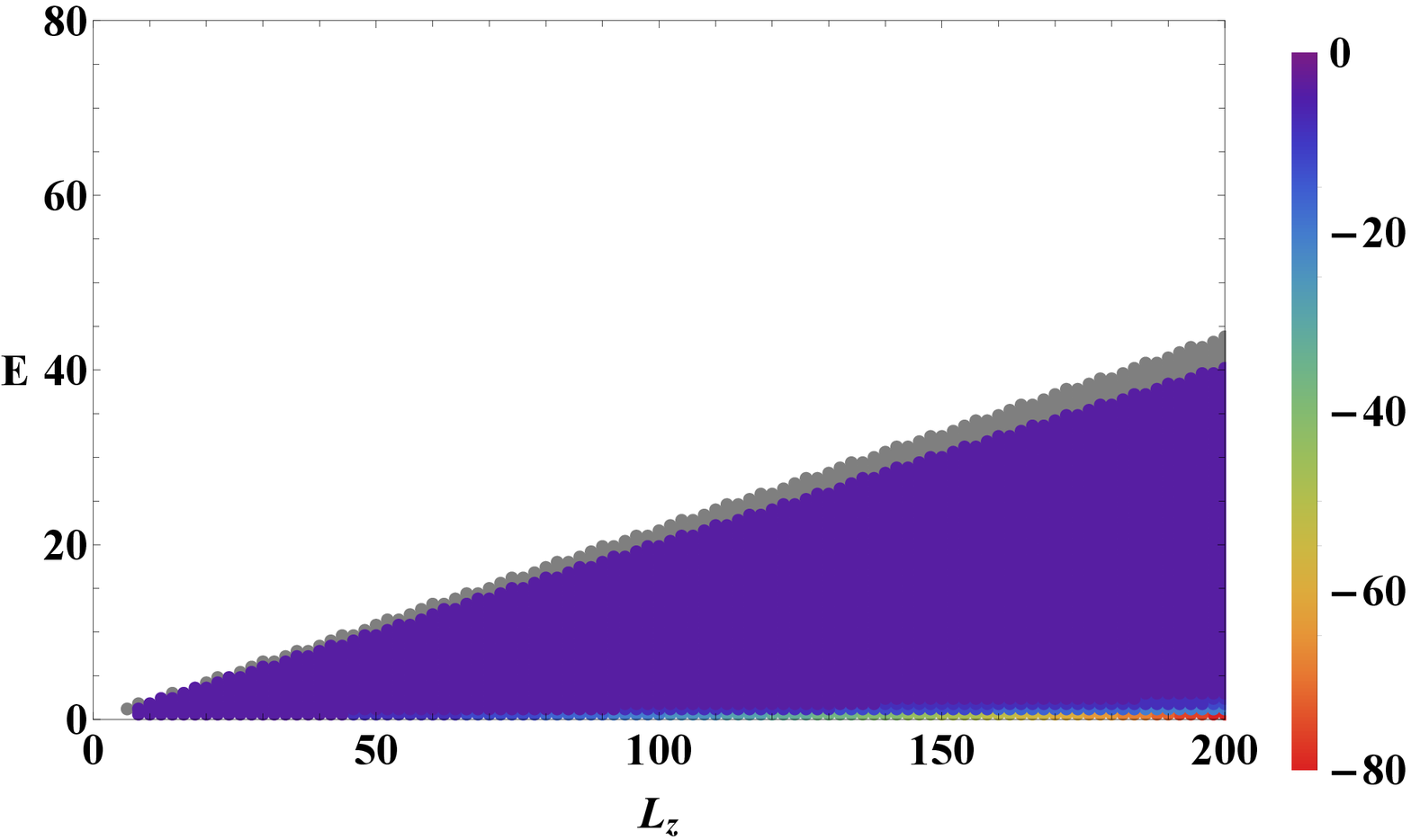}}
\subfigure[$w=3/2$ and $K = 0.8 K_{\rm{Max2}}$]
{\includegraphics[width=2. in]{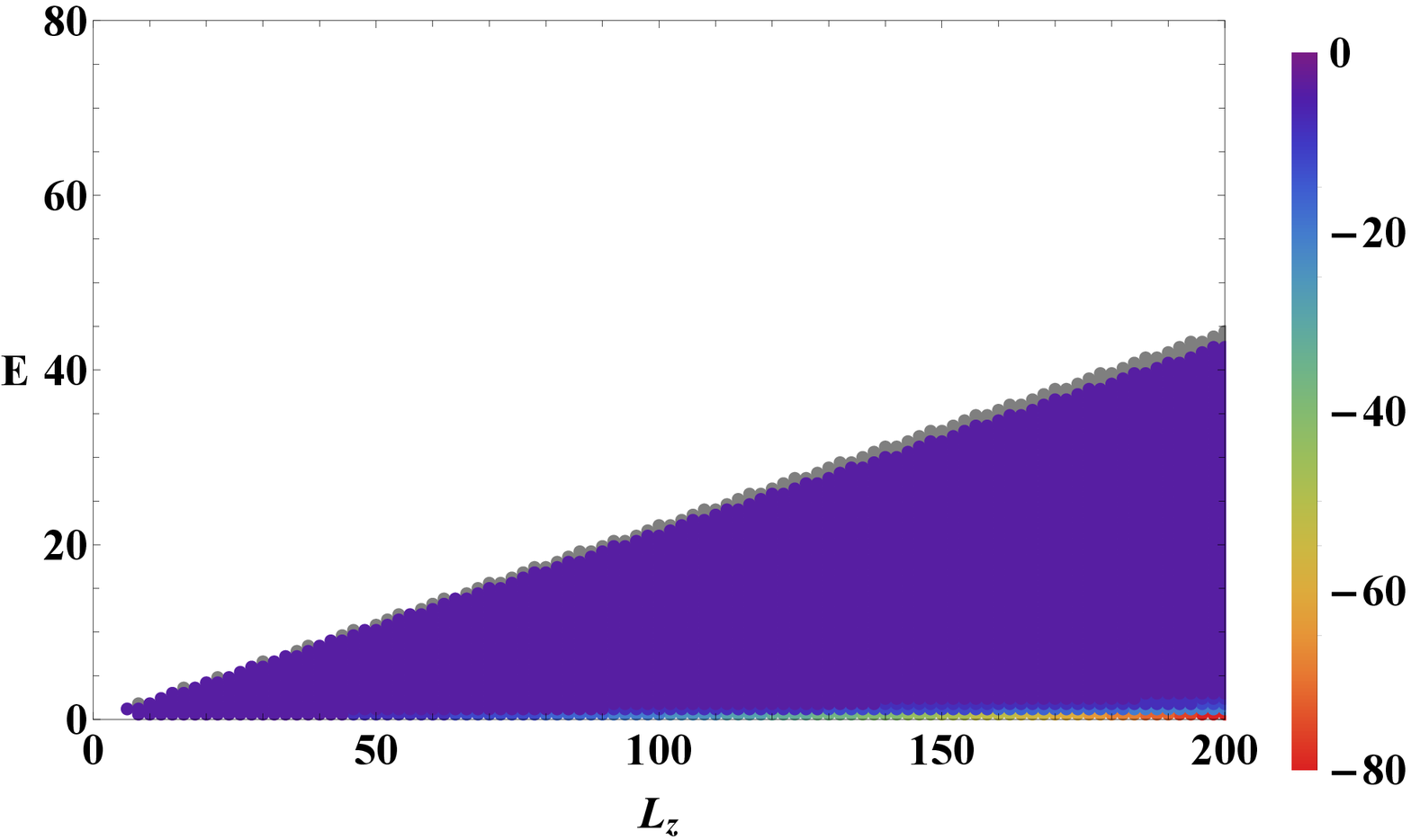}}
\subfigure[$w=3/2$ and $K = K_{\rm{Max2}}$]
{\includegraphics[width=2. in]{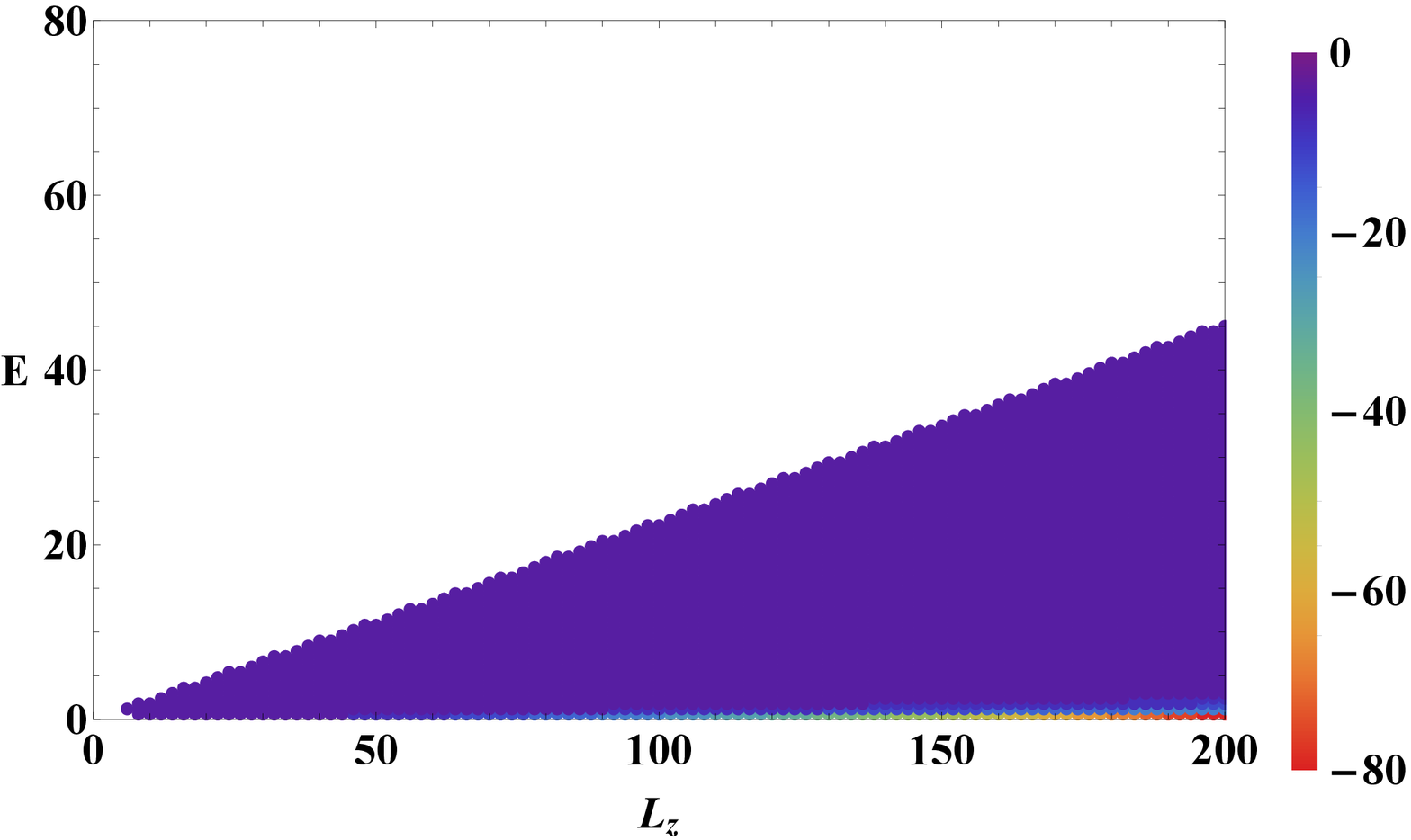}}
\subfigure[$w=2$ and $K = 0.6 K_{\rm{Max3}}$]
{\includegraphics[width=2. in]{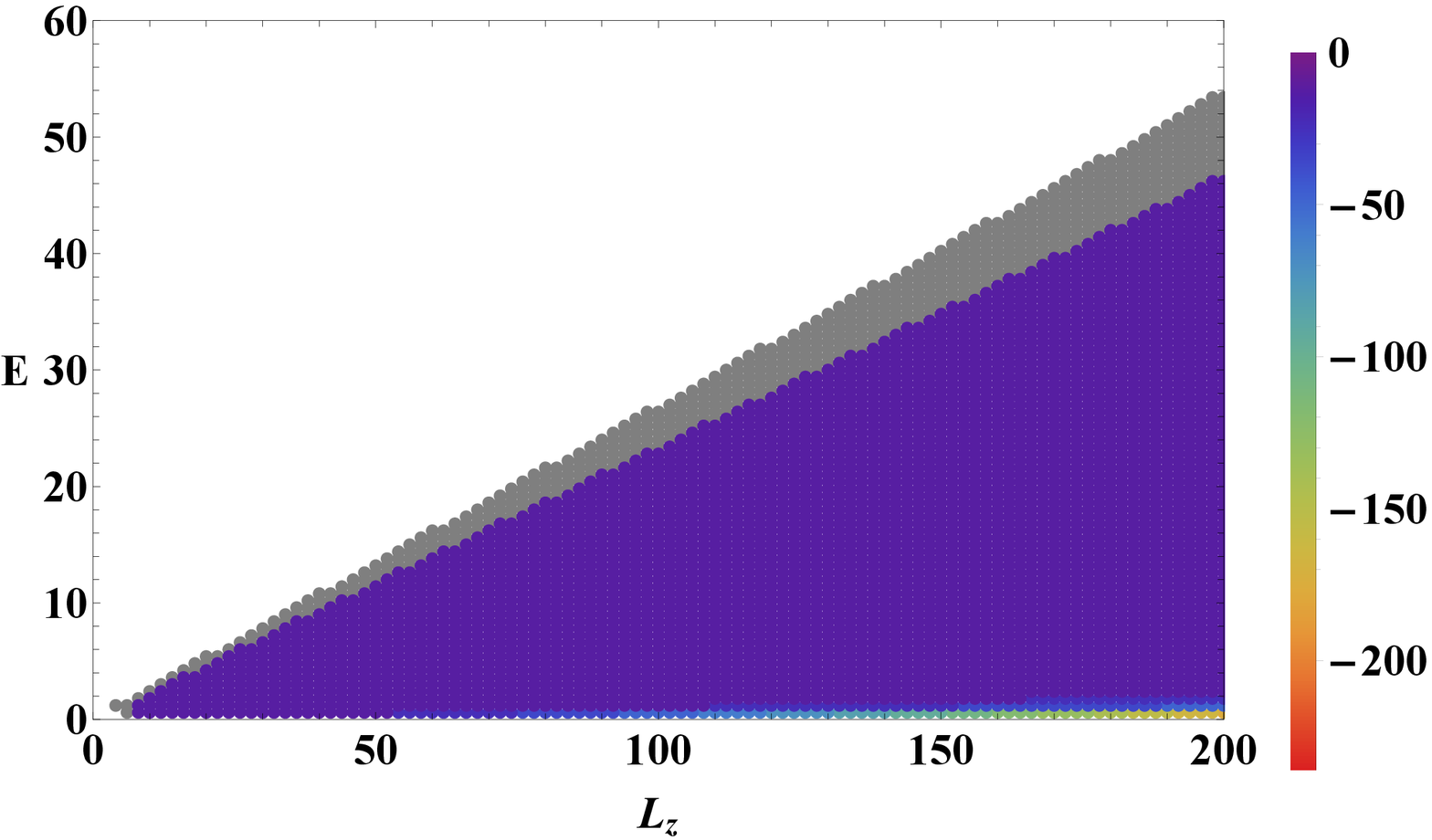}}
\subfigure[$w=2$ and $K = 0.8 K_{\rm{Max3}}$]
{\includegraphics[width=2. in]{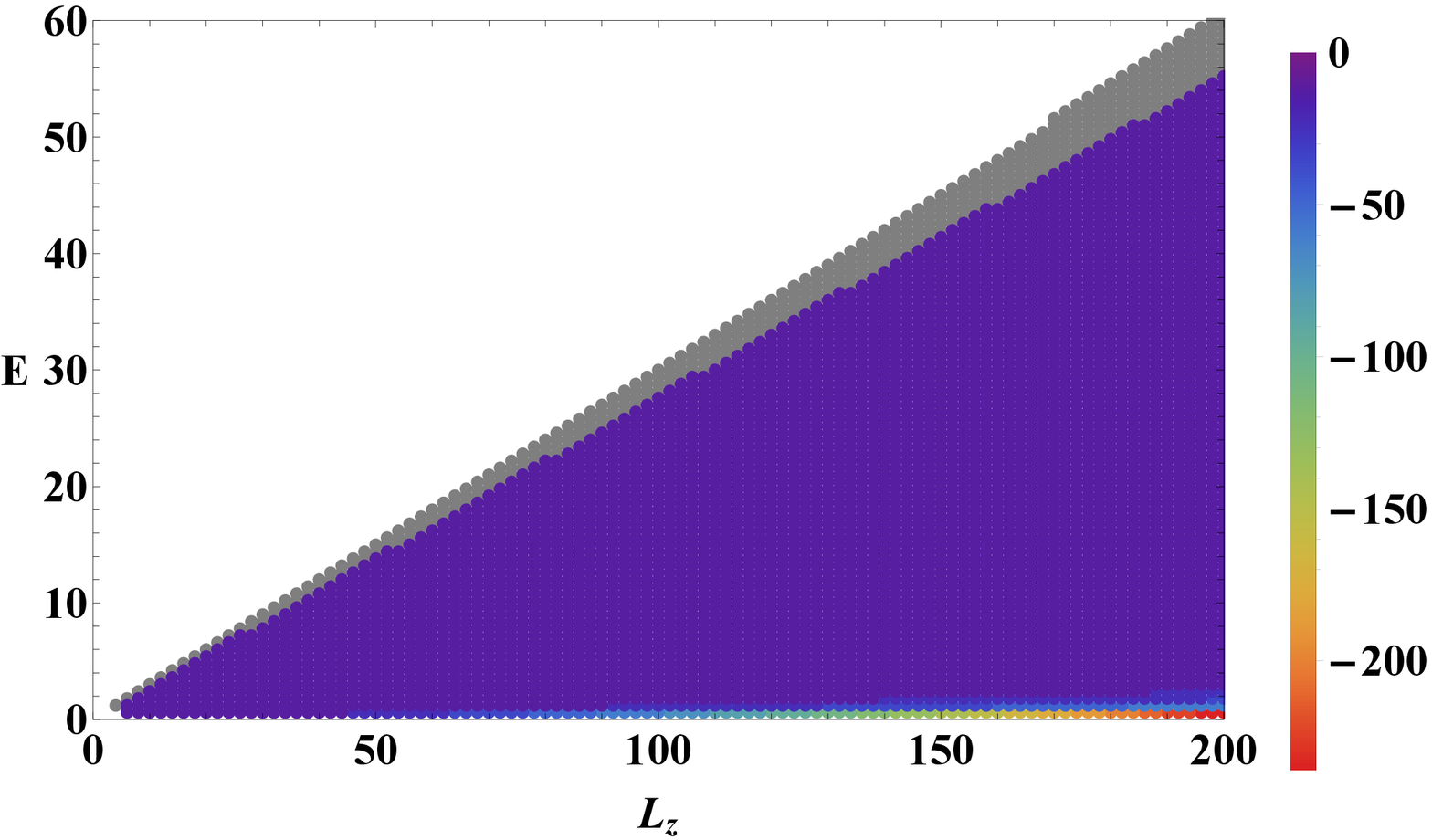}}
\subfigure[$w=2$ and $K = K_{\rm{Max3}}$]
{\includegraphics[width=2. in]{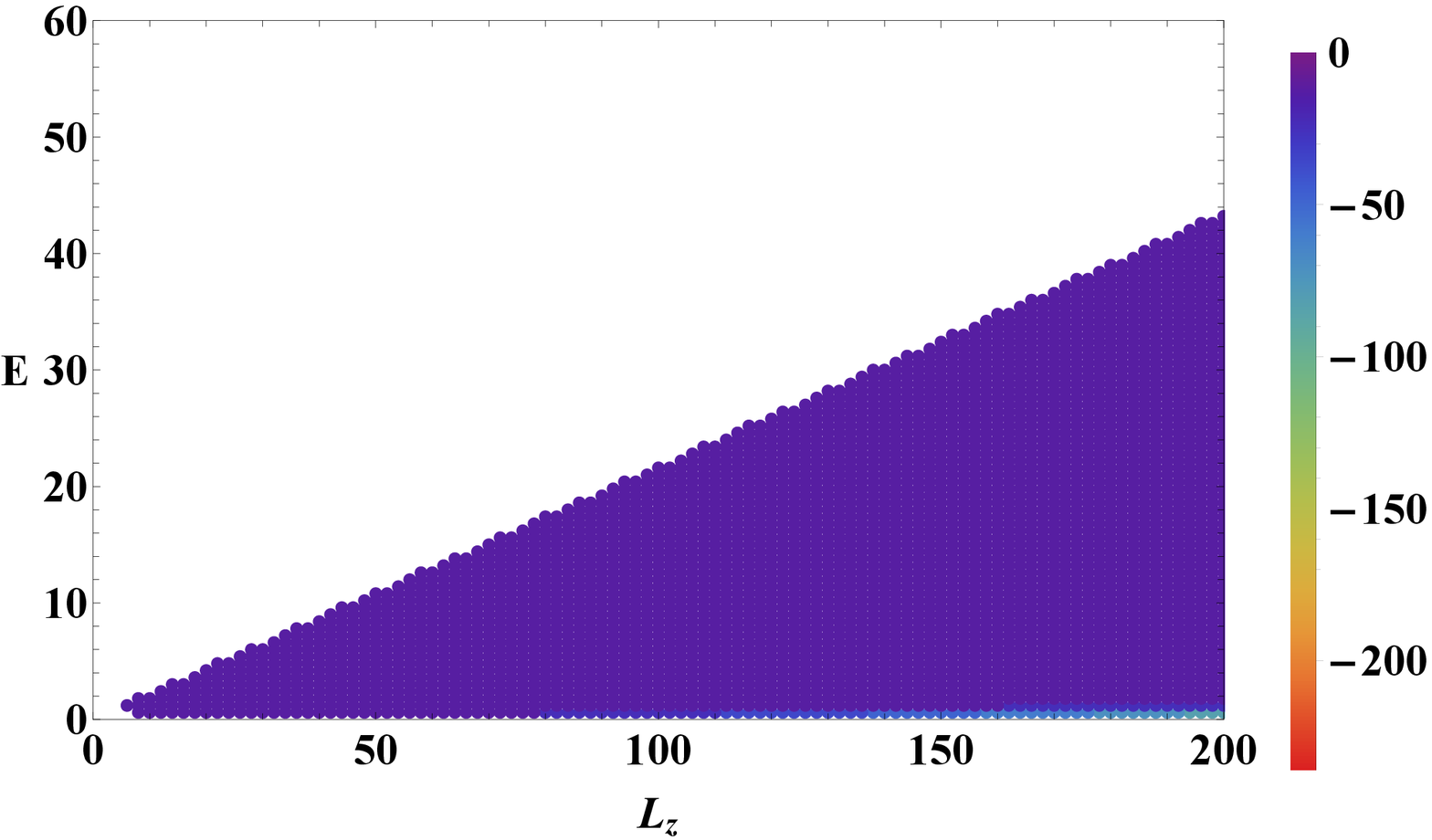}}
\end{center}
\caption{\footnotesize{(color online).
Violation of the Lyapunov bound with various $w$s.}}
\label{Lyapunov32}
\end{figure}

Figure \ref{Lyapunov32} represents the Lyapunov exponent, Eq.\ (\ref{Lyapunov2}).
Each row corresponds to $\omega = 2/3$, $\omega = 3/2$, and $\omega = 2$, respectively.
The right side of the color bar shows the value of $\kappa^2 - \lambda^2$.
The empty space in the upper left part of each figure corresponds to the region that does not satisfy ${\cal V}'_{\rm effc}=0$
or ${\cal V}''_{\rm effc}< 0$.
There are two types of points. The gray points mean the positive value and the colored points show the negative value.
The colored points point out the Lyapunov bound is violated and the redder colored point violates much bigger than the blue-colored point.
From the left side to the right side, the $K$ becomes larger and reaches to $K_{\rm Max}$,
which is consistent with the extremal black hole, and gray region vanishes,
then $\kappa^2 - \lambda^2$ is automatically negative.
As we expected, when $K$ grows to maximum $K_{\rm Max}$, the gray color turns into purple color.
We take $K_{\rm Max1}=-0.8375$, $K_{\rm Max2}=-0.8599$, and $K_{\rm Max3}=-1.1569$.
The absolute value of $\kappa^2 - \lambda^2$ is bigger with small energy and large angular momentum,
which is located at the bottom right as a red point.
We note that $\kappa^2$ is irrelevant to $E$, but $\lambda^2$ decreases when $E$ increases.
Therefore, we can estimate that the large $E$ will make it hard to violate the Lyapunov bound.

\section{Summary and discussions \label{sec4}}

\quad

We studied the homoclinic orbit of a particle in a black hole geometry coexisting
with anisotropic matters and whether its trajectory around the black hole could violate the well-known chaos bound.

It is known that the homoclinic orbit of particle motion corresponds to the boundary
between chaotic motion and nonchaotic motion.
We perturbed the particle with various values of energy and angular momentum at a position
near this homoclinic orbit and analyzed the Lyapunov exponent.

In Fig.\ \ref{Effective_potential}, we showed the effective potential and homoclinic orbit for
a test particle in the black hole geometry.
The existence of a local maximum in the effective potential is an important indicator
showing the existence of the homoclinic orbit. The orbit does not appear in Newtonian mechanics.
And also, it does not appear in the black hole geometry with naked singularity for type-$(ii)$.
It only appears in black holes that have the event horizon and with naked singularity for type-$(i)$.
The existence of a local maximum at the effective potential suggests that the Lyapunov exponent could be positive.
The black hole with naked singularity should be analyzed separately~\cite{Virbhadra:2002ju, Virbhadra:2007kw},
which will provide a testbed for the cosmic censorship hypothesis~\cite{Penrose:1969pc} to be proved by future detectors.

In Fig.\ \ref{Lyapunov30}, we analyzed $\lambda^2$ in terms of $L_z$ and $|K|$ at a given energy value.
Their behaviors show that the value of $\lambda^2$  increases when $L_z$ or $|K|$ increases.
In particular, the additional anisotropic matter field enhances the Lyapunov exponent.

In Fig.\ \ref{Lyapunov32}, we showed how the chaos bound $\kappa^2-\lambda^2$ varies in terms of
$E$ and $L_z$ with increasing $|K|$.
It can be seen that anisotropic matter could affect violate chaos bounds more.
We note that $\kappa^2$ is irrelevant to $E$, but $\lambda^2$ decreases when $E$ increases.
Therefore, we can estimate that the large $E$ will make it hard to violate the Lyapunov bound.

The trajectory of a particle around a black hole corresponds to classical motion.
On the other hand, the temperature of the black hole coming from the quantum effect
in the given geometry makes it a thermodynamic system. How can classical motion
and thermal effect (by the quantum effect) in the given geometry be related?
A relationship was conjectured in Ref.\ \cite{Maldacena:2015waa}.
It is an inequality equation in which the physical quantity
by the classical motion of particles in the spacetime geometry
should be bounded by the physical one due to the thermal effect.
It is known that one could define the black hole temperature by the surface gravity.
At the black hole event horizon, the surface gravity corresponds to the acceleration
as exerted at infinity necessary to keep the object at that horizon.
That acceleration decreases as a black hole approach the extremal one.
That is, in a thermodynamic system, the universal bound given to
the Lyapunov exponent by the motion of a classical particle will decrease and vanish.
In the black hole system, the analysis of the two physical quantities is no longer related
and has been separated. Thus it seems that one of two physical quantities could vanish
independently without relating to each other. As a result, the inequality bound can be violated.

In this paper, instead of Lagrangian formalism, we employed Hamilton-Jacobi formalism~\cite{Misner:1974qy}.
This one is indeed a powerful method to find general solutions to equations of motion.
Furthermore, it is possible to separate variables even if the number of cyclic variables is insufficient.
On top of that, the presence of more constants of motion emerges naturally.
In the Hamilton-Jacobi equation, we could carry out a separation of variables,
including four constants of motion, and as a result, the four equations of motion turned into first-order differential equations,
resulting in a completely integrable system.

We used the coordinate time $t$ instead of the proper time $\tau$ when calculating the Lyapunov exponent~\cite{Wu:2003pe}.
This was to see the increase in the distance with time between the particle of one homoclinic orbit
and another particle with different initial conditions.
Using time $t$, the effective potential in the radial direction becomes a slightly more complex function of $r$.
However, both the use of time $t$ and $\tau$ do not change the properties of the integrability.

One can explore two kinds of characteristics showing chaos phenomena. One corresponds to
the Lyapunov exponent, while the other to the Poincare section in the phase space.
We focused on examining the Lyapunov exponent in this paper.

In our case, we thought that the Poincare section might not show a chaotic behavior of the system.
Because the dynamical system we are dealing with is integrable~\cite{Frolov:2017kze},
i.e. it corresponds to a geodesic motion.
Compared with other papers that showed chaotic behavior through the Poincare section,
in our case, there exist enough constants of motion with that in the angle direction,
and those papers do not have the constant of motion in the angle direction.
In particular, the momentum does not commute with the Hamiltonian;
i.e.\ the system has a nontrivial effective potential with varying momentum in the angle direction.
Cases showing chaotic behavior in the Poincare section are
shown in the papers~\cite{Semerak:2010lzj, Semerak:2012dx, Hashimoto:2016dfz, Dalui:2018qqv}.

If we consider the black hole at the center of a galaxy and the actual particles moving around it,
the motion of particles occurring together with various physical phenomena near the black hole
does not correspond to the probe limit such as geodesic motion.
As a result, the particle motion should be different from the geodesic one,
and the surrounding matter should interact directly with the particle.
If we add those effects one by one, we will end up describing the motion of a particle moving in nature,
and that motion will not be integrable. From the viewpoint of analysis,
the number of constants of motion should be reduced by one at least.
Then in the Poincare section,
they will have nontrivial lines that show chaotic behavior~\cite{Sukova:2013jxa, Witzany:2015yqa, Polcar:2019wfi}.
The recent investigations of a black hole shadow involved analyzing the null geodesics~\cite{Younsi:2016azx, Cunha:2018acu, Badia:2020pnh, Lee:2021sws, Shaikh:2021yux, Roy:2021uye, Vagnozzi:2022moj, Zubair:2022fiw, Khodadi:2022ulo, Sau:2022afl}.
It would be interesting to see an analysis that includes chaotic phenomena.
We leave the study of particle motion with additional effects as future work.

\section*{Acknowledgments}
\quad B.-H.~Lee(support from Grant No. NRF-2020R1F1A1075472), H.~Lee, W.~Lee(support from Grant No. NRF-2022R1I1A1A01067336), and S.~Jeong were supported by the Basic Science Research Program of the National Research Foundation of Korea funded by the
Ministry of Education through Center for Quantum Spacetime (CQUeST) of Sogang University(Grant No. 2020R1A6A1A03047877).
We are grateful to Hyeong-Chan Kim and Youngone Lee for their hospitality during our visit to Korea National University of Transportation
and the 4th Tangeumdae Workshop, Inyong Cho to Seoul National University of Science and Technology, O-Kab Kwon to Sungkyunkwan University,
and Gungwon Kang, Kyung Kiu Kim, Jae-Weon Lee, Jungjai Lee, Chanyong Park, and Hyun Seok Yang for useful discussions.

\end{document}